\documentclass[conference]{IEEEtran}
\usepackage{cite}
\usepackage[cmex10]{amsmath}
\usepackage{algorithmic}
\usepackage{array}
\usepackage{amssymb}
\usepackage[table,xcdraw]{xcolor}
 \usepackage{multirow}
\usepackage{amsfonts, amscd, amsthm, xspace}
\usepackage{verbatim}
\usepackage{color,xcolor}
\usepackage{graphicx}
\usepackage{epstopdf}
\usepackage{subfigure}
\usepackage[font={small,it}]{caption}
\usepackage[T1]{fontenc}
\usepackage{url}
\usepackage{booktabs}
\newcommand{\subparagraph}{}
\newtheorem{definition}{Definition}
\newtheorem{theorem}{Theorem}

\newtheorem{lemma}{Lemma}

\newtheorem{example}{Example}

\newcommand{\upperRomannumeral}[1]{\uppercase\expandafter{\romannumeral#1}}

\newcommand{\C}{C} 
\newcommand{\cpri}{C^\prime} 
\newcommand{\chat}{\Hat{C}} 
\newcommand{\mhat}{\hat{m}}

\newcommand{\m}{m}

\newcommand{\M}{M} 

\newcommand{\mpri}{m^\prime} 
\newcommand{\X}{X} 
\newcommand{\randomX}{X_R} 
\newcommand{\randomMX}{X_R(m)} 

\newcommand{\Xpri}{X^\prime} 
\newcommand{\Y}{Y} 

\newcommand{\OneX}{\overrightarrow{X}_i} 

\newcommand{\OneY}{\overrightarrow{Y}_i}

\newcommand{\OneE}{\overrightarrow{E}_i} 
\newcommand{\R}{R} 

\newcommand{\Fq}{\mathbb{F}_q} 
\newcommand{\Ft}{\mathbb{F}_{2^t}} 

\newcommand{\n}{n} 
\newcommand{\key}{k} 
\newcommand{\prikey}{k^\prime}

\newcommand{\alllink}{L_1, L_2,  \cdots , L_C} 

\newcommand{\zi}{Z_R} 
\newcommand{\zo}{Z_W} 
\newcommand{\zio}{Z_{RW}} 
\newcommand{\zii}{Z_{RO}} 
\newcommand{\zoo}{Z_{WO}} 

\newcommand{\sizezi}{z_r} 
\newcommand{\sizezo}{z_w} 
\newcommand{\sizezio}{z_{rw}} 
\newcommand{\sizezii}{z_{ro}} 
\newcommand{\sizezoo}{z_{wo}} 

\newcommand{\sizezotilde}{\tilde{z}_w} 
\newcommand{\sizeziotildej}{\tilde{z}_{rw}^{(j)}} 
\newcommand{\sizezootildej}{\tilde{z}_{wo}^{(j)}} 
\newcommand{\badlinks}[1]{Z_w^{(i_{#1})}}
\newcommand{\goodlinks}[1]{\bar{Z}_w^{(i_{#1})}}
\newcommand{\GF}[1]{\mathbb{F}_{#1}}

\newcommand{\Li}{L_i}

\newcommand{\hide}[1]{}
\newcommand{\Hp}[1]{H\left( #1 \right)}
\newcommand{\Hcond}[2]{H\left( {#1} | {#2} \right)}
\newcommand{\I}[2]{I\left( {#1} ; {#2} \right)}

\newcommand{\K}{K} 
\newcommand{\twomatrix}[2]{\begin{bmatrix} #1 \\ #2\end{bmatrix}}

\newcommand{\Xsub}[1]{\overrightarrow{X}_{#1}}
\newcommand{\Rjsadd}{R_{j,s}^{add}}
\newcommand{\Rjsow}{R_{j,s}^{ow}}
\newcommand{\Rjaddc}{R_{j}^{add}}
\newcommand{\Rjowc}{R_{j}^{ow}}
\newcommand{\Rjfadd}{R_{j,f}^{add}} 
\newcommand{\Rjfow}{R_{j,f}^{ow}} 
\newcommand{\zaddfb}{\mathcal{Z}_{pos,fb}^{add}} 
\newcommand{\Xone}{\X^{(1)}} 
\newcommand{\Yone}{\Y^{(1)}} 

\newcommand{\vecX}{\overrightarrow{X}}
\newcommand{\vecU}{\overrightarrow{U}}
\newcommand{\OneU}{\overrightarrow{U}_i} 
\newcommand{\SecondU}{\overrightarrow{U}_j} 
\newcommand{\primex}{[\OneU \OneK \vec{h_{i1}}, \vec{h_{i2}}, ... , \vec{h_{i\C}}]} 
\newcommand{\OneK}{\overrightarrow{K}_i}
\newcommand{\SecondK}{\overrightarrow{K}_j}
\newcommand{\primey}{[\overrightarrow{U}_i^{\prime} \: \vec{K_{i}}^{\prime T}\:  \vec{h}_{i1}^{\prime T} \: \vec{h}_{i2}^{\prime T} \cdots  \vec{h}_{iC}^{\prime T}]}

\newcommand{\xone}{x^{(1)}} 
\newcommand{\xtwo}{x^{(2)}} 
\newcommand{\yone}{y^{(1)}} 
\newcommand{\ytwo}{y^{(2)}} 
\newcommand{\Zfalse}{Z_{\times}} 
\newcommand{\Ztrue}{Z_{\checkmark}} 
\newcommand{\Zhatfalse}{\hat{Z}_{\times}} 
\newcommand{\Zhattrue}{\hat{Z}_{\checkmark}} 

\newcommand{\epsn}{\epsilon_n}

\newcommand{\bb}{b} 
\newcommand{\N}{N} 
\newcommand{\vechij}{\vec{h_{ij}}}
\newcommand{\zbar}{\vec{z}}

\newcommand{\addpr}{\mathcal{Z}_{w, nf}^{add}}
\newcommand{\owpr}{\mathcal{Z}_{w, nf}^{ow}}

\newcommand\blfootnote[1]{%
  \begingroup
  \renewcommand\thefootnote{}\footnote{#1}%
  \addtocounter{footnote}{-1}%
  \endgroup
}

\begin{document}

\title{Reliable and secret communication over \\ adversarial multi-path networks}

\author{Qiaosheng (Eric) Zhang${}^{*1}$\qquad\qquad Swanand Kadhe${}^{*2}$\vspace{0.5em}\\ 
Mayank Bakshi${}^{1}$\qquad\qquad Sidharth Jaggi${}^{1}$\qquad\qquad Alex Sprintson${}^2$\vspace{0.3em}\\ \small ${}^1$Institute of Network Coding, Chinese University of Hong Kong \qquad${}^2$Texas A\& M University}

\maketitle

\blfootnote{The work of Qiaosheng (Eric) Zhang, Mayank Bakshi and Sidharth Jaggi was partially supported by a grant from University Grants Committee of the Hong Kong Special Administrative Region, China (Project No. AoE/E-02/08).
The work of Alex Sprintson was partially supported by the NSF under grant CNS-0954153 and by the AFOSR under contract No. FA9550-13-1-0008. 

* indicates equal contribution. }

\IEEEpeerreviewmaketitle

\begin{abstract}

We consider the problem of communication over a \emph{multi-path} network in the presence of a causal adversary. The limited-view causal adversary is able to eavesdrop on a subset of links and also jam on a potentially overlapping subset of links based on the current and past information. To ensure that the communication takes place reliably and secretly, resilient network codes with necessary redundancy are needed.  We study two adversarial models -- \emph{additive} and \emph{overwrite jamming} and we optionally assume {\em passive} feedback from decoder to encoder, {\em i.e.}, the encoder sees everything that the decoder sees. The problem assumes transmissions are in the {\it large alphabet} regime. For both jamming models, we find the capacity under four scenarios -- reliability without feedback, reliability and secrecy without feedback, reliability with passive feedback, reliability and secrecy with passive feedback. We observe that, in comparison to the non-causal setting, the capacity with a causal adversary is strictly increased for a wide variety of parameter settings and present our intuition through several examples. 
\end{abstract}

\section{Introduction}
\label{sec:introduction}

Consider the following example of a communication problem. Alice wishes to wirelessly transmit a message $m$ to receiver Bob by communicating over $C$ different frequencies. Their communication is intercepted by a limited-view adversary Calvin who has his receiver tuned to subset $Z_R$ of the frequencies, and can jam a potentially overlapping subset $Z_W$ of frequencies by adding transmissions on them. Due to the nature of the channel, Calvin can only see the signal up to the current time to maliciously determine his jamming strategy for the current time instant. We wish to answer questions of the following form: ``\emph{Without knowing which frequencies Calvin is monitoring/jamming, what is the maximum communication rate at which Bob can decode Alice's message successfully, while keeping the message secret from Calvin?}''. This example corresponds to a model in which Alice wishes to communicate reliably and secretly with Bob over a channel with an eavesdropper/additive jammer. A variant of the problem is when, additionally, Alice can also hear the channel outputs (she too is monitoring all $C$ frequencies, and therefore has passive feedback). In this variant we wish to understand whether this knowledge can improve the best possible rate.

We model this problem as that of communication over a noiseless multi-path network consisting of $\C$ parallel links between the sender and the receiver. As mentioned above, the adversary Calvin can eavesdrop on a subset ${\zi}$ and jam on a subset ${\zo}$. Subsets ${\zio}$ , ${\zii}$ and ${\zoo}$\footnote{RW, RO, WO stand for ``read and write'', ``read only'' and ``write only'' respectively. In a wireless setting these sets may correspond to different physical constraints on Calvin's ability to eavesdrop on or jam certain frequencies. In distributed storage system setting these sets may correspond to Calvin having or read-or-write, read-only, or write-only permissions on different devices.} represent the links that Calvin can both eavesdrop on and jam, only eavesdrop on (but not jam) and only jam (but not eavesdrop on) respectively. In addition, the sizes of ${\zio}$, ${\zii}$ and ${\zoo}$ are bounded from ${\sizezio}$, ${\sizezii}$ and ${\sizezoo}$. The adversarial vector ${\zbar}$ = (${\sizezio}, {\sizezii}, {\sizezoo}$) measures Calvin's power. Moreover, Calvin also knows the encoding and decoding schemes so that he may mimic Alice's behavior to confuse Bob. We consider a {\em causal} constraint on Calvin's behaviors, {\em i.e.}, Calvin can only use the knowledge of symbols up to the current time slot to decide his jamming strategy. 

\subsection*{Related work}
\textbf{Reliable communication:} The problem of reliable communication (with no secrecy constraints) against a malicious eavesdropping adversary has been well-studied in the past. The maximum possible rate has been characterized under various settings -- both causal and non-causal. The non-causal setting is relatively well understood both in the classical error-correction setup~\cite{ReedS:60,Singleton:64,Ozarow.Wyner1985,LapidothN:98} and the network error correction setting ~\cite{JagLHE:05,Yao.Silva.Jaggi.Langberg2010:Secure-Error-Correcting,JaggiL:12,zhang2015talking}. A key feature of these results is that in many of these models, Calvin can decrease the capacity by \emph{twice} the number of links he controls, by ``pushing'' Alice's transmissions towards the ``nearest plausible transmission'', thereby inflicting ``double- damage''. This heuristic also suggests an intuitive scheme for Bob's decoder -- try to detect as many corrupted links as possible and treat those as erasures -- in this case Calvin's actions would only cause ``single damage''. Critically, Calvin's ability to cause double damage depends on his ability to be able to see the full transmission for each link in $Z_R$ before determining the optimal jamming strategy for $Z_W$.

In contrast, in the causal setting, by using stochastic encoding, the adversary may not be able to predict some of the future symbols, which can then be used to detect the set $\zo$. Causal adversaries for classical channel coding and network coding problems have also been well studied~\cite{dey2013codes, Nutman.Langberg2008}. In \cite{kosut2014generalized}, the authors consider causal, omniscient adversaries for multicast networks, and characterize precise conditions under which any positive rate is achievable. Further, they also provide some upper and lower bounds on the rates. However, the question of characterizing the capacity in general networks containing malicious jammers remains open in the main.

\textbf{Reliable and secret communication:} The problem of \emph{both} reliable \emph{and} secure communication over a network has also received considerable attention in the literature. For reliable, secure communication, \cite{Jag:05} characterize the capacity when $z_{ro} = z_{wo} = 0$. In \cite{Yao.Silva.Jaggi.Langberg2010:Secure-Error-Correcting}, the authors consider another extreme when the set of edges that are eavesdropped and jammed are disjoint, \emph{i.e.}, $z_{rw} = 0$. The capacity for a general $\vec{z} = (z_{rw}, z_{ro}, z_{wo})$ for a non-causal adversary with no feedback has been considered in a previous work \cite{zhang2015talking} by the authors of this work.

Another model that is related to our setup is that of an adversarial wiretap (AWTP) channel \cite{wang2014efficient}, wherein the adversary can eavesdrop up to a given fraction of symbols sent over a channel, and can jam another (possibly intersecting) fraction of symbols based on what he eavesdrops. There are two key differences from our work: (i) the authors only consider the problem of additive jamming, and (ii) the capacity characterization is parametrized with ``coarser granularity'', in that instead of parametrizing the problem in terms of ($z_{rw}$, $z_{ro}$, $z_{wo}$), the authors parametrize it in terms of ($z_{rw} + z_{ro}, z_{rw} + z_{wo}$), \emph{i.e.}, in terms of the total number of eavesdropped and jammed links.

Another problem that is closely related to ours is that of Secure Message Transmission (SMT) \cite{DolCWY:93, patra2010unconditionally}. Under SMT, a sender aims to communicate a message reliably and secretly to the receiver over multiple parallel links out of which a fraction of links are eavesdropped and another (possibly intersecting) fraction are jammed. There are a several differences from our model: (i) The SMT problem focusses on computing a lower bound on the number of links that are required for reliable and secret communication of one message symbol, and usually do not provide information-theoretically tight capacity characterizations, (ii) most schemes are multi-round, 2-way protocols where the receiver can (actively) talk to the sender (though some protocols are indeed 1-way), (iii) the problem parametrization is again in terms of ($z_{rw} + z_{ro}, z_{rw} + z_{wo}$).

\subsection*{Our contributions}
We consider the problem of causal jamming with an optional secrecy requirement. Taking cue from our prior work~\cite{zhang2015talking}, we consider a finer characterization of the adversary's power by classifying his controlled links into read-only, write-only, and read-and-write subsets. We examine this problem in two settings --  additive and overwrite jamming. The motivation for an additive adversary comes from wireless networks, where, the adversary may add his own signal to the transmitted signal. On the other hand, the overwrite adversary models the adversarial action in a wired network, where the adversary is more likely to completely replace the true transmitted packets with fake packets of his choice. \footnote{Notice that for a write only link, the overwrite adversary knows the output codewords while the additive adversary has no way to learn the output codewords.}

In the setting without feedback, Theorems~\ref{thm:additiveequal}-\ref{thm:overwriteunequal} state the capacity when secrecy is not required. Theorems~\ref{thm:secret-reliable-rate-additive} and~\ref{thm:secret-reliable-rate-overwrite} state our results with secrecy requirement and also without feedback. When passive feedback is available to the encoder, we characterize the capacity in Theorems~\ref{thm:ajcf} and~\ref{thm:ojcf}, and also consider secrecy in Theorems 9 and 10. 
 
The rest of this paper is organized as follows.  In Section~\ref{sec:example}, we illustrate the role of causality in limiting the adversary's power by giving four examples. We formally define the problem in Section~\ref{sec:statement} and state our main results with short proof sketches in Section~\ref{sec:result}. The detailed proofs are presented in Appendices~\ref{sec:lemma}-\ref{sec:appendix_feedback_secrecy}.

\section{Key ideas}
\label{sec:example}
The techniques used in this paper build upon the ideas introduced in~\cite{Jag:05, zhang2015talking}. In this section, we present a short intuitive overview of some of these ideas via toy examples. 
\subsection{Ideas for achievability}
\subsubsection{Reed-Solomon codes} The application of Reed-Solomon codes~\cite{ReedS:60} (or in fact, any MDS code) to network error-correction is well-studied. These are particularly useful when the parameters $\sizezio,\sizezii$, and $\sizezoo$  correspond to a ``strong adversary'' that can choose both the set $\zo$ and the corrupted codewords in a ``worst-case'' manner.\footnote{The simplest example of a strong adversary is an omniscient one, {\em i.e.}, when $\sizezi=\C$. However, the exact parameter settings that correspond to a strong adversary depend on the flavour of the problem being considered, {\em e.g.} causal vs non-causal, overwrite vs additive etc.} If Bob were able to detect the set $\zo$ with high probability, this would allow him to treat the set $\zo$ as ``erasures'', thus enabling a rate of $\C-\sizezo$ to be achievable -- indeed, this is what some of the schemes we present attempt to do.\footnote{It is important here to make a distinction between zero error probability ({\em i.e.}, robustness to worst-case errors) and vanishing error probability requirements. In the former case one  can use the Singleton bound~\cite{Singleton:64} to see that the best achievable rate is $\C-2\sizezo$ regardless of what the adversary knows. However, the Singleton bound requires Calvin to know the entire transmission. Hence, in the latter setup, a higher rate may be achievable if the adversary has limited eavesdropping power.}

\subsubsection{Pairwise-hashing} When the adversary has limited knowledge of the transmitted codewords, in some settings a {\em pairwise-hashing} scheme is useful in detecting the set $\zo$ and enabling treating the corrupted set of links as erasures~\cite{Jag:05}.The following example presents the main intuition here. 
\begin{example}[\em Limited-view non-causal adversary~\cite{zhang2015talking}]\label{ex:pairwise} \normalfont Consider a network with three links $L_1,L_2$ and $L_3$, and an adversary Calvin that can both read and write on exactly one link, {\em i.e.}, $\sizezio=1$ and $\sizezoo=\sizezii=0$. Even though the zero-error capacity of this network is $\C-2\sizezo=1$, we argue that in the vanishing error probability setup, the  link $\zio$ can be detected and the rate is $\C-\sizezio=2$. The codeword sent on the link $L_i$ consists of three parts 
-- the $i$-th symbol from a $(3,2)$ Reed-Solomon codeword $U_i$, a random key $K_i$, and hash values $H_{i1}, H_{i2}, H_{i3}$ with $H_{ij}=h(K_i,U_j)$ for a suitably designed non-linear hash function $h(\cdot,\cdot)$. Upon receiving the codewords, Bob checks consistency within each pair of links $(L_i,L_j)$ by verifying if the received values satisfy both $H_{ij}=h(K_i,U_j)$ and $H_{ji}=h(K_j,U_i)$. Without loss of generality, assume that Calvin choses $\zio=\{L_1\}$ (unknown to Alice and Bob) and corrupts $(U_1,K_1,H_{11},H_{12},H_{13})$. Note that since Calvin does not know the values of $K_2$ and $K_3$, it can be shown that the probability that he can satisfy the checks for $H_{21}$ and $H_{31}$ is small if he choses to change $U_1$.  On the other hand, links $L_2$ and $L_3$ cross-verify all of their mutual hashes successfully, thus isolating $L_1$ as the corrupted link. As a result, Bob can successfully decode the message by treating $U_1$ as an erasure for the Reed-Solomon code.
\qed
\end{example}
The scheme in the Example~\ref{ex:pairwise} exploits the adversary knowing only a subset of what the decoder sees. In the problems considered in this paper, the additional assumption of causality further limits his view and enables a higher rate. For example, in the three link network above, even if we permit the adversary to read all the links ({\em i.e.}, $\sizezii=2,\sizezio=1$), the fact that the adversary does not know the random keys $K_1,K_2,K_3$ while perturbing the selected $U_i$ prevent him from being able to deterministically match the corresponding pairwise hashes. This allows the decoder to detect the corrupted link. 

\subsubsection{Adversary detection with passive feedback}
We use a similar idea as above in the setting with passive feedback ({\em i.e.}, the encoder overhears all past symbols received by the decoder before transmitting the current symbol). In this case, the rate $\C-\sizezo$ is possible for an even larger set of parameters by using a multi-round scheme that lets Bob detect the set of corrupted links.
\begin{example}[\em Adversary detection] \label{ex:passivefeedback}{\normalfont
Consider a two link network with $\sizezio=1,\sizezii=\sizezoo=0$. Here, no positive rate is possible without feedback ({\em c.f.} Example~\ref{ex:zero}). However, with passive feedback, a rate of $1$ is possible. To see this, consider a two-round scheme. Alice first generates two independent and uniformly random keys $K_1$ and $K_2$. Next, for a message $M$, Alice transmits $(M,K_1)$ and $(\M,K_2)$ on links $L_1$ and $L_2$ respectively in the first round. Now, Alice sees both the received codewords, say $(\hat{\M}_1,\hat{K}_1)$ and $(\hat{\M}_2,\hat{K}_2)$, and determines if any of the links were corrupted by Calvin. In the second round, Alice sends hash $H=(h(\hat{\M}_1,\hat{K}_1),h(\hat{\M}_2,\hat{K}_2))$ on all links that she determines to be uncorrupted in the first round, while sending uniformly random bits having the same length as $H$ on the corrupted links. This lets Bob detect and ignore any link corrupted in the first round and decode the message from the uncorrupted link(s). Note that since Calvin sees only one link, he cannot ensure that the second round codeword on any link corrupted by him satisfies the hash equation.}   
\end{example}

The intuition is that the encoder can use his feedback to first determine which links have been corrupted by the adversary and then convey this information to the decoder by sending values consistent with the hash function only on the uncorrupted links (and inconsistent values on others). 
\subsubsection{Mixing keys for secrecy} In the setting where information theoretic secrecy is demanded in addition to reliability, we use standard one-time pad arguments that mix the message with random keys. Since the adversary can see up to $\sizezi$ links, as long as the key rate is at least $\sizezi$, we show that the secrecy requirement is met. The error correcting code is chosen so that both the message and the key are decoded by the receiver. As a result, as long as the overall rate decreases by $\sizezi$, both secrecy and reliability are simultaneously met. \footnote{Note that decreasing the rate by $\sizezi$ suffices even when Alice does not know the set of links corrupted by Calvin. Surprisingly, when Alice can learn the set of links corrupted by Calvin in previous rounds of a multi-round scheme, the secrecy capacity may be higher (as seen in Example~\ref{ex:ske})}
\subsubsection{Secret key extraction via passive feedback}
A surprising effect of passive feedback is that it allows Alice and Bob to extract secret keys from corrupted links as long as the received codeword on the link is hidden from Calvin. This allows for simultaneous reliable and secret transmission at rates higher than that achievable by just mixing $\sizezi$ random keys with the message. The following example of an {\em additive} adversary ({\em i.e.}, on $\zoo$ links, Calvin adds an error vector to the transmitted codeword without necessarily knowing what was sent by Alice) illustrates the key idea that enables achieving the rate $\min\{C-\sizezi,C-\sizezo\}$ in the additive setting.
\begin{example}[\em Secret key extraction]\label{ex:ske} {\normalfont
Consider a two link network with $\sizezii=\sizezoo=1$. Here, no positive rate is possible for simultaneous reliable and secret transmission without feedback. However, with passive feedback, the following multi-round protocol achieves an asymptotic rate of $1$. Let the message $M=M_1M_2\ldots M_n$ be a length-$n$ binary vector. The transmission is divided into three-stages. In the $i$-th round of the first stage, Alice generates a random bit $K_j$ and sends $X_{1,j}=M_j$ on the first link and $X_{2,j}=M_j\oplus K_j$ on the second link. After each bit is received, Alice checks if any of the two links have been corrupted by Calvin. Let $c$ be the first round where Calvin has corrupted one of the links, say link $L_i$, {\em i.e.}, $Y_{i,c}=X_{i,c}\oplus E_c$ for some $E_c$. Alice now starts the second stage of transmission. In each round $j=c+1,\ldots,n+1$, Alice sends $X_{i,j}=K_j$ on the corrupted link $L_i$ and $X_{i\oplus 1+1,j}=M_{j-1}+Y_{1,j-1}$ on the uncorrupted link. Note that since Calvin has revealed that $L_i\in\zoo$, Alice knows that Calvin cannot infer the values of $Y_{i,j}$ as each $X_{i,j}$ is independent of $X_{i\oplus 1+1,1},\ldots,X_{i\oplus 1+1,n+1}$. Thus, in the second stage of transmission, $Y_{i,j}$ acts as a shared key for secret transmission on the uncorrupted link for the $(i+1)$-th round. Finally, in the third stage, Alice transmits the values of $c$ and $i$ by using the reliable transmission scheme (without secrecy) of Example~\ref{ex:passivefeedback}.}
\end{example}
\subsection{Ideas for the converse}

\subsubsection{Cut-set bound} Since the adversary can corrupt $\sizezo$ links, he can replace the codewords on the links in $\zo$ by uniformly random symbols independent of the original codewords. By a simple argument based on Fano's inequality, one can conclude that no rate higher than $\C-\sizezo$ is possible if the error probability must vanish to $0$.
\subsubsection{A symmetrization argument\cite{dey2013codes}} A tighter converse than the above can be obtained when the adversary is ``powerful enough''. For example, in the setting without feedback, if the adversary can corrupt at least half of the links, no positive rate is possible. The argument here is similar to the Singleton Bound~\cite{Singleton:64} and allows for stochastic encoding as well. The following example illustrates the idea.

\begin{example}[\em Symmetrization with a causal adversary]\label{ex:zero} \normalfont Consider a three link network under attack from a causal  adversary with $\sizezio=2$ and $\sizezoo=\sizezii=0$. It is straightforward to see that the pairwise hashing scheme can be foiled by the adversary -- he can ensure that the two links in $\zo$ satisfy each other's hashes and thus there is no reliable way for the decoder to determine the uncorrupted link. More generally, for any coding scheme, the adversary can follow a {\em symmetrization} strategy as follows. Suppose the encoder maps message $m$ to codewords $x_1,x_2,x_3$ according to an encoder conditional probability distribution $p_{X_1,X_2,X_3|M}$.\footnote{Note that both a deterministic encoder as well as an encoder using pairwise hashing can be viewed as special cases under this formalism.} The adversary first chooses a message $m'$ uniformly at random from the set of possible messages and draws codewords $x_1',x_2',x_3'$ according to the conditional probability distribution $p_{X_1,X_2,X_3|M}\left(x_1',x_2',x_3'|m'\right)$. Next, the adversary chooses $\zo$ to be either $\{L_1\}$ or $\{L_2,L_3\}$ with equal probability and replaces the codewords $(x_i:i\in\zo)$ by $(x_i':i\in\zo)$. Now, from the decoder's point of view, given the received codewords $y_1,y_2,y_3$, the messages $m$ and $m'$ are equally likely. Thus, the decoder cannot reliably determine whether the true message is $m$ or $m'$ and as a result the error probability is bounded away from $0$ for any code of positive rate.\qed
\end{example}
When feedback is present, an argument similar to the above holds, albeit for a smaller set of adversarial parameters. With feedback, since the code operates over multiple rounds, the adversary needs to be sufficiently powerful to be able to fool the decoder even when the encoder knows the received codewords on the links in $\zo$.

\section{Problem Statement}
\label{sec:statement}

\begin{figure}[!t]
	\begin{center}
	\includegraphics[scale=0.45]{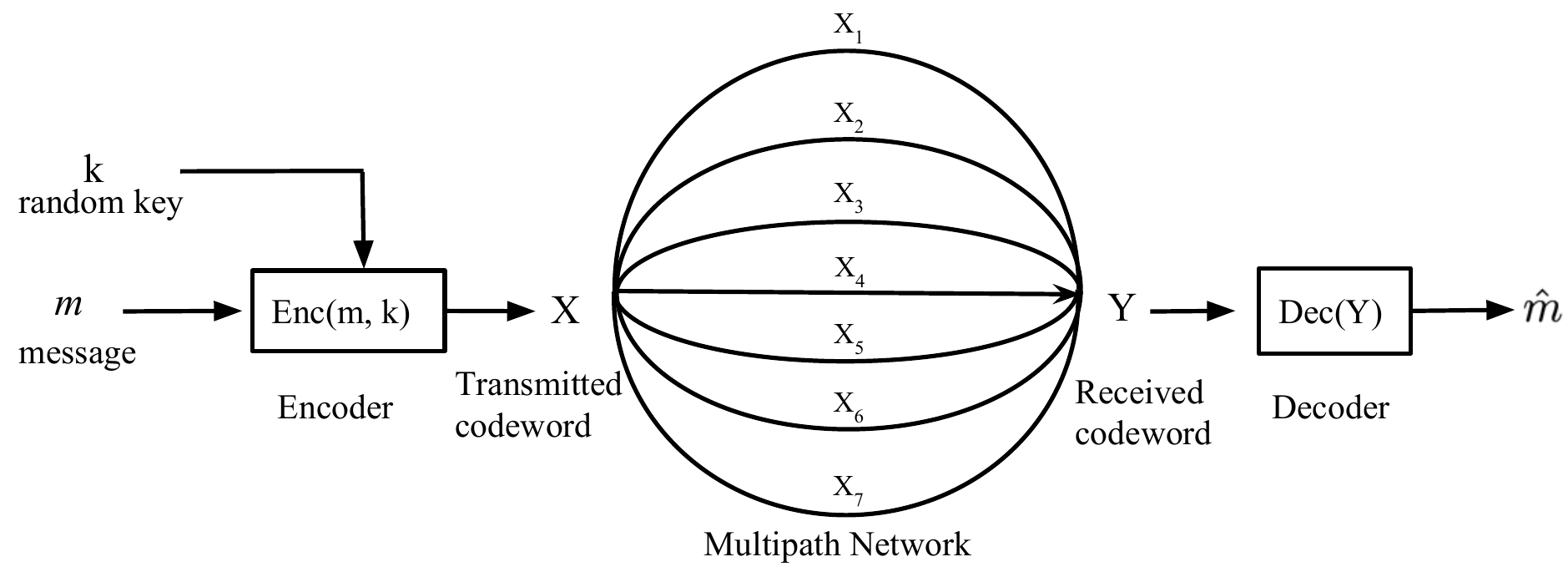}
	\caption{\underline{System diagram} for a multi-path network consisting of $\C$ parallel links ($\C = 7$ in this example). }
	\label{fig:system}
	\end{center}
\end{figure}

The multi-path network model and the encoding/decoding process are illustrated in Figure~\ref{fig:system}. In our setting, the multi-path network consists of $\C$ parallel, directed links $L_1$, $L_2$,\ldots,$L_C$. For the equal link capacity network, each link is of unit capacity. For a general unequal link capacity network, the capacity of link $L_i$ is denoted by $u_i$, $\forall i \in \left\{ 0, 1,\ldots,\C\right\}$. The total capacity of a unequal link capacity network is defined as $\chat$ bits per use, where ${\chat} := \sum\limits_{i=1}^{C}{u_i}$. In particular, the total capacity of the equal link capacity is $\C$ bits per use. We also optionally consider {\em passive feedback} available causally at the transmitter, {\em i.e.}, the transmitted overhears the received symbols at the decoder causally. 

We begin by formally describing the encoder, decoder, and possible adversarial actions for a code for  \emph{block length} $\n$, \emph{i.e.}, for a code that operates over $\n$ time steps and $r$ rounds ($r\leq n$ with $r=1$  denoting the case without feedback). For simplicity, we focus on networks with equal link capacity and deal with the unequal link capacity case only informally. 

  In the following, matrices $\X=[\X^{(1)}\X^{(2)}\ldots \X^{(r)}]$ and $\Y=[\Y^{(1)}\Y^{(2)}\ldots \Y^{(r)}]$ respectively denote the collection of transmitted and received codewords across all links for all rounds. Here, $\X^{(k)}$ (resp. $\Y^{(k)}$) is a matrix whose $i$-th row $\vec{X}_i^{(k)}$ (resp. $\vec{Y}_i^{(k)}$) denotes the transmitted (resp. received) codeword on link $L_i$ in the $k$-th round. 

\subsection{Encoder}
The transmitter Alice encodes a $\n\R$-bit message $\m$, where $\R$ stands for the message rate. The message $\m$ is assumed to be uniformly distributed from $\left\{ 0, 1\right\}^{{\n}{\R}}$. To perform a stochastic encoding, a random key $\key$, which is uniformed distributed from the finite field ${\Ft}$, is also generated by Alice. 

In the first round, Alice encodes $\m$ into a collection of $C$ $n$-length codewords $\vec{X}_1^{(1)},\vec{X}_2^{(1)},\ldots,\vec{X}_\C^{(1)}$. In this case, the encoder function for the first round takes the form $$\Psi_e^{(1)}:\left\{ 0, 1\right\}^{\R\n} \times {\Ft} \rightarrow \left\{ 0, 1\right\}^{\C\n^{(1)}}.$$ If no feedback is present, no further actions are performed by Alice. In this case, $n^{(1)}=n$. 

If feedback is present, in each subsequent round $k$, Alice's codewords  $\vec{X}_1^{(k)}, \vec{X}_2^{(k)},\ldots,\vec{X}_\C^{(k)}$  are each of length $n^{(k)}$ and are determined by the message $\m$, the random key $\key$, and the feedback from prior rounds $\Y^{(1)}\ldots\Y^{(k-1)}$. Formally, Alice's encoder for the $i$-th round takes the form $$\Psi_e^{(k)}:\left\{ 0, 1\right\}^{\R\n} \times {\Ft} \times\prod_{j=1}^{k-1}\left\{ 0, 1\right\}^{\C\n^{(j)}} \rightarrow \left\{ 0, 1\right\}^{\C\n^{(k)}},$$ where, $\sum_{k=1}^r\n^{(k)}=n$.

\subsection{Decoder}

The decoder Bob receives the code matrix ${\Y}$, which may be different from $\X$ and outputs a reconstruction $\mhat$. The decoding function takes the form $$\gamma_e({\Y}):  \left\{ 0, 1\right\}^{{\n}{\C}} \rightarrow \left\{ 0, 1\right\}^{{\n}{\R}}.$$ 

\subsection{Adversary}

Out of the $\C$ links of the multi-path network, the adversary Calvin is able to eavesdrop (but not jam) a subset $\zii$ of size $\sizezii$, jam (but not eavesdrop) a subset $\zoo$ of size $\sizezoo$, both eavesdrop and jam a subset $\zio$ of $\sizezio$. Calvin's power is measured by the adversarial vector ${\zbar}$ = (${\sizezio}, {\sizezii}, {\sizezoo}$). The encoding and decoding strategies are known to Calvin. However, the two end users do not know how Calvin chooses $\zii$, $\zoo$ and $\zio$ in advance.

\subsubsection{Additive and Overwrite Jamming}

An additive jammer may induce additive bias on the transmitted codeword. Assume the codeword transmitted on $\Li$ is $\OneX$ and the bias is $\OneE$, the received codeword would be $\OneY = \OneX + \OneE$. 
On the other hand, an overwrite adversary can overwrite the transmitted codeword by its own one directly. If the codeword is $\OneX$ and the bias is $\OneE$, the received codeword will be $\OneY = \OneE$.

\subsubsection{Causal Adversary}

We restrict the adversary to be {\em causal}, {\em i.e.}, the adversary is only allowed to jam the symbol of current time slot based on the observation of current and past time slots. More specifically, at any given time $t$, given a $\C\times\n$ code matrix $\X$, the adversary can use the knowledge of only the first $t$ symbols from rows in subset $\zio\cup\zii$ in order to jam the $t$-th symbols from the rows in $\zio\cup\zoo$.  

In contrast, a non-causal adversary~\cite{longversion} enjoys the full knowledge of all the symbols in the rows belonging to $\zio\cup\zii$ at all times. Obviously, the non-causal adversary has a stronger power and leads to a potentially lower rate.

\subsubsection{Reliability and Security}

Instead of a zero-error probability, we aim to achieve an $\varepsilon$-error probability. The communication is reliable if for any $\varepsilon > 0$, by choosing $\n$ large enough, there exists a code of block length $\n$ such that the error probability $P_e = Pr[m \ne \mhat] < \varepsilon$.

In terms of security, we aim to achieve the information-theoretically perfect secrecy. Assume the subset Calvin can eavesdrop is $\zi = \{i_1, i_2, \cdots, i_{\sizezi}\}$ and the sub-codeword on $\zi$ links is $\X_{\zi} = [\Xsub{i_1}^T \: \Xsub{i_2}^T \cdots \Xsub{i_{\sizezi}}^T]^T$. To achieve security, the mutual-information between the message and Calvin's observation should be zero, i.e. $\I{\M}{\X_{\zi}} = 0$.

\section{Main Results}
\label{sec:result}
In this section, we present the main results and sketch their proofs. The full proofs of the results can be found in Appendices~\ref{sec:proof_causal}-\ref{sec:appendix_feedback_secrecy}. We group our results into four parts -- reliability without feedback, reliability and secrecy without feedback, reliability with feedback, as well as reliability and secrecy with feedback. For each of these cases, we discuss the additive jamming and the overwrite jamming separately. Finally, we give a ``complete'' characterization of the problem in Table \ref{my_table}.

\subsection{Reliability without feedback}
For both additive and overwrite jammers, we obtain a two-part rate-region, i.e. \emph{weak adversary regime} and \emph{strong adversary regime}. The weak adversary regime for additive jamming is 
\begin{IEEEeqnarray*}{rCl}
\label{eq:weak-add}
{\addpr} &=& \left\{ \vec{z}: z_{wo} + 2z_{rw} < C\right\},
\end{IEEEeqnarray*}
whereas for overwrite jamming, the weak adversary regime is  
\begin{IEEEeqnarray*}{rCl}
\label{eq:weak-ow}
{\owpr} &=& \left\{ \vec{z}: 2z_{wo} + 2z_{rw} < C\right\}.
\end{IEEEeqnarray*}
The strong adversary regime equals the complement of the weak adversary regime. In the weak adversary regime, the achievability relies on erasure codes coupled with the pairwise-hashing scheme. The rate is limited to zero in the strong adversary regime. 

First we consider the scenario when reliable communication is the only objective. For equal and unequal link capacity networks, for any $\zbar = ({\sizezio}, {\sizezii}, {\sizezoo})$ such that ${\sizezio} + {\sizezii} + {\sizezoo} \le {\C}$, the maximum achievable reliable rates for additive and overwrite jamming are characterized in the following. 

\begin{theorem}[Additive jamming for equal link capacities]\label{thm:additiveequal} 
Under additive causal jamming, the maximum achievable reliable rate is 
\begin{equation*}
  \Rjaddc(\C, \zbar) = 
      \begin{cases}
      C - (z_{rw} + z_{wo}), & \ \mbox{if} \ \vec{z} \in \addpr, \\
      0, & \ \mbox{otherwise}.
      \end{cases}
  \label{eq:thm-1}
\end{equation*}
\end{theorem}

\begin{theorem}[Overwrite jamming for equal link capacities]\label{thm:overwriteequal}
Under overwrite causal jamming, the maximum achievable reliable rate is 
\begin{equation*}
  \Rjowc(\C, \zbar) = 
      \begin{cases}
      C - (z_{rw} + z_{wo}), & \ \mbox{if} \ \vec{z} \in \owpr, \\
      0, & \ \mbox{otherwise}.
      \end{cases}
  \label{eq:thm-2}
\end{equation*}
\end{theorem}

For additive jamming and overwrite jamming, the two-part rate-regions are slightly different though the expressions are similar. Note that, regardless of the coding scheme, the best rate that we can hope for is $\C - \sizezo$ since the adversary can corrupt any $\sizezo$ links. In the weak adversary regime, the best rate $\C - \sizezo$ is indeed achievable. To achieve it, the encoder would use an erasure code to encode the message and apply the pairwise-hashing scheme to help detect errors. The decoder detects the corrupted links first (which are regarded as erasures) and then the message will be retrieved from the sub-codewords carried by the uncorrupted links. 

However, no coding scheme (including pairwise-hashing) works for the strong adversary regime. The adversary can always adopt a ``symmetrization'' strategy so that the decoder is unable to distinguish the correct message and the fake message. The proof of the converse relies on an argument based on the Singleton bound~\cite{Singleton:64} that we present in Appendix~\ref{sec:proof_causal}. 
 
\begin{figure}[t]
	\begin{center}
	\includegraphics[scale=0.37]{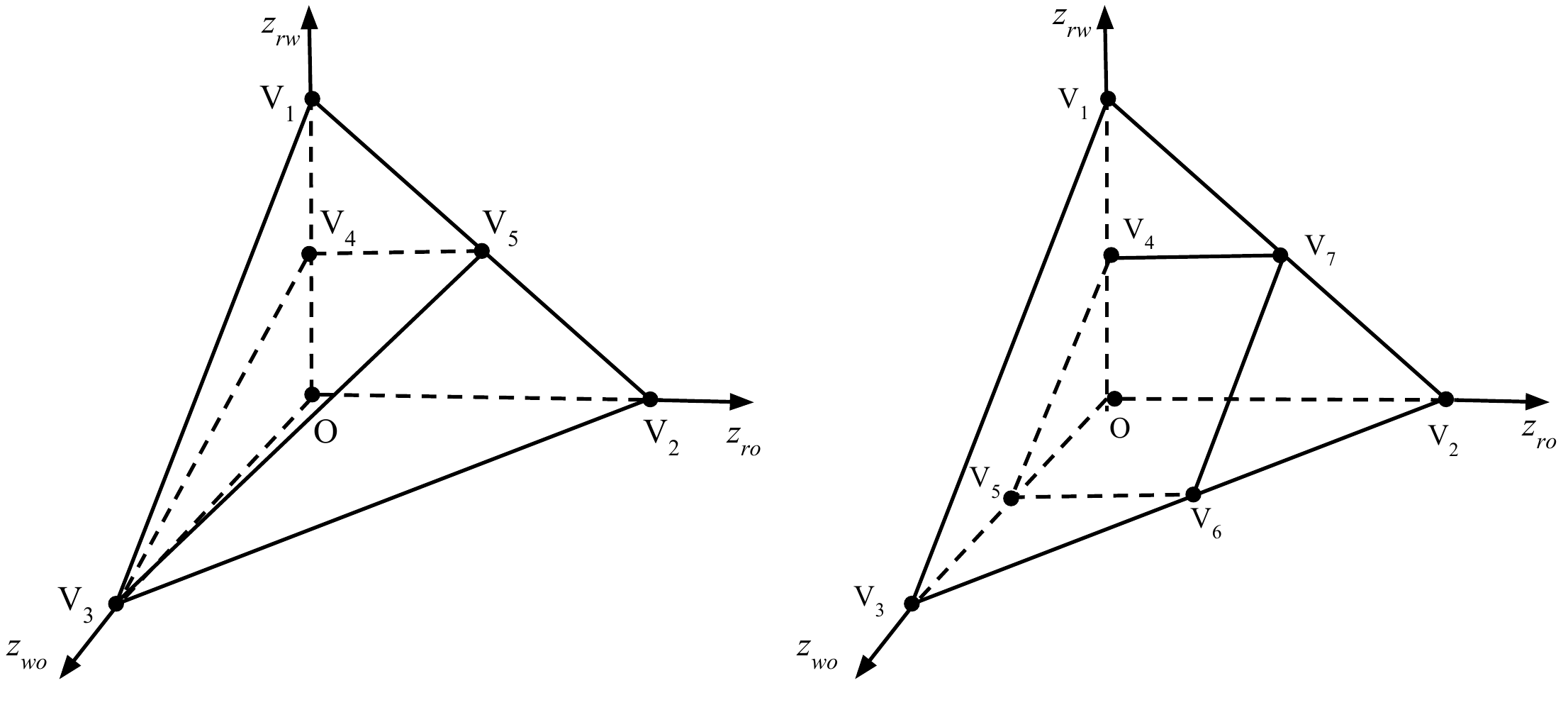}
	\caption{The rate regions for additive and overwrite causal jamming adversaries (Theorem 1 and 2): for additive (resp. overwrite) jamming, the pentahedron $V_3OV_2V_5V_4$ (resp. $OV_4V_5V_2V_7V_6$) represents the \emph{weak adversary} regime, while the polyhedron $V_3V_4V_5V_1$ (resp. $V_1V_3V_5V_4V_7V_6$) represents the \emph{strong adversary} regime.} 
	\label{fig:3d}
	\end{center}
\end{figure}

{\em \underline{Unequal link capacities}:} To incur maximum damage, the adversary may choose links with highest sum-rate to attack. We define the the total capacity of any subset of size $w$ links as $U_w$. Different choice of the subsets may incur different values of $U_w$. Typically, the notation $(U_w)_{max}$ is used to denote the maximum value of $U_w$, i.e. the largest sum-capacity of all possible subsets of size $w$. Similar to the situation with equal link capacities, the main idea is also encoding by erasure codes as well as adopting ``pairwise-hashing'' scheme to detect adversarial attacks. However, the only difference is that the maximum rate depends on Calvin's ability to corrupt the links with largest sum-capacities.

\begin{theorem}[Additive jamming for unequal link capacities]\label{thm:additiveunequal}
Under additive causal jamming, the maximum achievable reliable rate is 
\begin{equation*}
  \Rjaddc(\chat, \zbar) = 
      \begin{cases}
      \Hat{C} - (U_{(z_{rw} + z_{wo})})_{max}, & \ \mbox{if} \ \vec{z} \in \addpr \\
      0, & \ \mbox{otherwise}.
      \end{cases}
  \label{eq:thm-3}
\end{equation*}
\end{theorem}

\begin{theorem}[Overwrite jamming and unequal link capacities]\label{thm:overwriteunequal}
Under overwrite causal jamming, the maximum achievable reliable rate is 
\begin{equation*}
  \Rjowc(\chat, \zbar) = 
      \begin{cases}
      \Hat{C} - (U_{(z_{rw} + z_{wo})})_{max}, & \ \mbox{if} \ \vec{z} \in \owpr, \\
      0, & \ \mbox{otherwise}.
      \end{cases}
  \label{eq:thm-4}
\end{equation*}
\end{theorem}

\subsection{Reliability and secrecy without feedback}
\label{sec:secrecy}
In this section, we consider the scenario wherein Calvin tries to learn some information about Alice's message from the links he eavesdrops. Besides reliable communication, we also want to prevent Calvin from gaining any information about the message. For this, we consider information-theoretically perfect secrecy, which requires that $\I{\M}{\X_{\zi}} = 0$, where $\X_{\zi}$ is the sub-codeword transmitted on the links in $\zi$. In the following, we characterize the reliable and secure rate region in the equal link capacity case for the causal adversary.

\begin{theorem}[Additive, causal jamming with secrecy, equal link capacities]
\label{thm:secret-reliable-rate-additive}
Under additive causal jamming, the maximum achievable reliable and secret rate is 
\begin{equation*}
\label{eq:secret-reliable-rate-additive}
\Rjsadd(\zbar) = 
\begin{cases}
\left(C - (\sizezoo + \sizezii + 2\sizezio)\right)^{+} & \text{if} \ \vec{z} \in \addpr,\\
0 & \text{otherwise},
\end{cases}
\end{equation*}
where $(x)^{+}$ is defined as $(x)^{+} = \max\left\{0, x\right\}$.
\end{theorem}

\begin{theorem}[Overwrite, causal jamming with secrecy, equal link capacities]
\label{thm:secret-reliable-rate-overwrite}
Under overwrite causal jamming, the maximum achievable reliable and secret rate is 
\begin{equation*}
\label{eq:secret-reliable-rate-owerwrite}
\Rjsow(\zbar) = 
\begin{cases}
\left(C - (\sizezoo + \sizezii + 2\sizezio)\right)^{+} & \text{if} \ \vec{z} \in \owpr,\\
0 & \text{otherwise},
\end{cases}
\end{equation*}
where $(x)^{+}$ is defined as $(x)^{+} = \max\left\{0, x\right\}$.
\end{theorem}

The converse for the weak adversary regime (for both additive and overwrite jamming) follows from the standard information-theoretic inequalities, where we use the secrecy condition that any subset of $\sizezi = (\sizezio+\sizezii)$ links cannot carry any meaningful information. In the achievable scheme, Alice needs to {\it mix} her message with $\sizezi$ random keys and then use the reliable encoding scheme consisting of pairwise hashing and erasure coding. For the strong adversary regime, the converse is based on the Singleton-type arguments similar to the only reliability case. We present the detail proof in Appendix~\ref{sec:proofs-secrecy-additive}. 

\begin{table*}[t]
\scriptsize
\centering
\caption{The table gives expressions for the information-theoretically optimal rate regions in each scenario. The yellow cells refer to the main results presented in \cite{zhang2015talking}. The new results presented in this paper are shown in red cells. We use $(x)^{+}$ to represent $\max\left\{0, x\right\}$.}
\label{my_table}
\begin{tabular}{|c|c|l|l|l|}
\hline
\multicolumn{2}{|c|}{Model}                                                                                    
& \multicolumn{1}{c|}{regime} 
& \multicolumn{1}{c|}{reliability} 
& \multicolumn{1}{c|}{reliability \& secrecy}\\ 
\hline
\multirow{4}{*}{\begin{tabular}[c]{@{}c@{}}Non-\\ causal\end{tabular}}            
& \multirow{2}{*}{additive}  
&\cellcolor[HTML]{FFFFC7}$z_{ro}+z_{wo}+2z_{rw} < C$   & \cellcolor[HTML]{FFFFC7}\begin{tabular}[c]{@{}l@{}}$C - (z_{rw} + z_{wo})$ \\ $\hat{C} - (\Lambda_{z_{rw} + z_{wo}})_{max}$\end{tabular}  
& \cellcolor[HTML]{FFFFC7}$\left(C - z_{ro} - z_{wo} - 2z_{rw}\right)^{+}$ \\ \cline{3-5} 
& {}
&  \cellcolor[HTML]{FFFFC7}$z_{ro}+z_{wo}+2z_{rw} \ge C$ 
&  \cellcolor[HTML]{FFFFC7}\begin{tabular}[c]{@{}l@{}}$(C - 2z_{rw} - z_{wo})^+$ \\ $(\hat{C} - (\Lambda_{2z_{rw} + z_{wo}})_{max})^+$\end{tabular} & \cellcolor[HTML]{FFFFC7}$0$   \\ \cline{2-5} 
& \multirow{2}{*}{overwrite} 
& \cellcolor[HTML]{FFFFC7}$z_{ro}+2z_{wo}+2z_{rw} < C$ 
& \cellcolor[HTML]{FFFFC7}\begin{tabular}[c]{@{}l@{}}$C - (z_{rw} + z_{wo})$ \\ $\hat{C} - (\Lambda_{z_{rw} + z_{wo}})_{max}$\end{tabular} 
& \cellcolor[HTML]{FFFFC7}$\left(C - z_{ro} - z_{wo} - 2z_{rw}\right)^{+}$  \\ \cline{3-5} 
&{}
& \cellcolor[HTML]{FFFFC7}$z_{ro}+2z_{wo}+2z_{rw} \ge C$  
& \cellcolor[HTML]{FFFFC7}\begin{tabular}[c]{@{}l@{}}$(C - 2z_{rw} - 2z_{wo})^+$ \\ $(\hat{C} - (\Lambda_{2z_{rw} + 2z_{wo}})_{max})^+$\end{tabular}  
& \cellcolor[HTML]{FFFFC7}$0$   \\ 
\hline
\multirow{4}{*}{\begin{tabular}[c]{@{}c@{}}Causal w/o\\ feedback\end{tabular}} 
& \multirow{2}{*}{additive}  &\cellcolor[HTML]{F9EEED} $z_{wo}+2z_{rw} < C$  
& \cellcolor[HTML]{F9EEED}\begin{tabular}[c]{@{}l@{}}$C - (z_{rw} + z_{wo})$ \\ $\hat{C} - (\Lambda_{z_{rw} + z_{wo}})_{max}$\end{tabular}  
&  \cellcolor[HTML]{F9EEED}$\left(C - z_{ro} - z_{wo} - 2z_{rw}\right)^{+}$   \\ \cline{3-5} 
&{}
& \cellcolor[HTML]{F9EEED}$z_{wo}+2z_{rw} \ge C$  
& \cellcolor[HTML]{F9EEED}$0$ 
& \cellcolor[HTML]{F9EEED}$0$       \\ \cline{2-5} 
& \multirow{2}{*}{overwrite} 
& \cellcolor[HTML]{F9EEED}$2z_{wo}+2z_{rw} < C$  
& \cellcolor[HTML]{F9EEED}\begin{tabular}[c]{@{}l@{}}$C - (z_{rw} + z_{wo})$ \\ $\hat{C} - (\Lambda_{z_{rw} + z_{wo}})_{max}$\end{tabular} 
&  \cellcolor[HTML]{F9EEED}$\left(C - z_{ro} - z_{wo} - 2z_{rw}\right)^{+}$    \\ \cline{3-5} 
&{}
& \cellcolor[HTML]{F9EEED}$z_{wo}+2z_{rw} \ge C$   
& \cellcolor[HTML]{F9EEED}$0$    
& \cellcolor[HTML]{F9EEED}$0$     \\ 
\hline
\multirow{4}{*}{\begin{tabular}[c]{@{}c@{}}Causal with \\ passive feedback\end{tabular}}     
& \multirow{2}{*}{additive}  
& \cellcolor[HTML]{F9EEED}$\left\{z_r = C \ \mbox{and} \ 2z_w < C\right\}\bigcup \left\{z_r < C\right\}$    
& \cellcolor[HTML]{F9EEED}$C - (z_{rw} + z_{wo})$
&  \cellcolor[HTML]{F9EEED}$\min\left\{C - z_r, C - z_w\right\}$     \\ \cline{3-5} 
&
{}
& \cellcolor[HTML]{F9EEED}$z_r = C \ \mbox{and}\ 2z_w \ge C$   
& \cellcolor[HTML]{F9EEED}$0$   
& \cellcolor[HTML]{F9EEED}$0$   \\ \cline{2-5} 
& \multirow{2}{*}{overwrite} 
& \cellcolor[HTML]{F9EEED}$z_{ro} + z_{wo} + z_{rw} < C$  
& \cellcolor[HTML]{F9EEED}$C - (z_{rw} + z_{wo})$ 
& \cellcolor[HTML]{F9EEED}$(C - z_{ro} - z_{wo} - z_{rw})^+$ \\ \cline{3-5} 
& 
& \cellcolor[HTML]{F9EEED}$z_{ro} + z_{wo} + z_{rw} = C$ 
& \cellcolor[HTML]{F9EEED}$\left(C - 2z_{wo} - 2z_{rw}\right)^{+}$  
& \cellcolor[HTML]{F9EEED}$0$ \\ 
\hline
\end{tabular}
\end{table*}

\subsection{Reliability with passive feedback}
In this section, we examine the effect of passive feedback on the capacity under jamming for both additive and overwrite settings. For both these cases, the parameter space again decomposes into two parts - the \emph{weak adversary regime} and \emph{strong adversary regime}. 

The main idea for achievability of the claimed rate in the weak adversary regime is to use a two-round code. The first round involves sending a code that can handle up to $\sizezio+\sizezoo$ erasures. At the end of the first round, Alice sees the codewords received by Bob and determines the links which have been corrupted. In the next round, Alice sends a random hash of all the received codewords by Bob on the uncorrupted links from first round. Bob can then check the received values from the second round to determine the links where the hash values do not match the received codeword and treat those links as erasures. 

The above scheme works as long as there is at least one link whose output is not seen by Calvin. This corresponds exactly to the condition for the weak adversary in the following theorems. If Calvin is able to see the output of all the links, he is as powerful as Alice and feedback no longer helps.

\begin{theorem}[Additive Jamming with Causal Feedback]\label{thm:ajcf} Under an additive jamming with causal feedback, the capacity is 
$$\Rjfadd=\left\{\begin{array}{ll}0& \mbox{if }\sizezi=\C\mbox{ and }\C\leq2\sizezo\\
C-\sizezo&\mbox{otherwise}\end{array}\right.$$
\end{theorem}

\begin{theorem}[Overwrite Jamming with Causal Feedback]\label{thm:ojcf} Under an overwrite jamming with causal feedback, the capacity is 
$$\Rjfow=\left\{\begin{array}{ll}\max{\{\C-2\sizezo,0\}}& \mbox{if }\sizezii+\sizezio+\sizezoo=\C\\
C-\sizezo&\mbox{otherwise}\end{array}\right.$$
\end{theorem}

We present the detailed proof in Appendix~\ref{sec:proof_of_feedback}.

\subsection{Reliability and secrecy with passive feedback}

The schemes for secrecy are similar to that for only reliability. The idea here is to mix appropriate number of random keys so as to prevent Calvin from inferring meaningful information. The protocols operate over multiple rounds. For overwrite jamming, Alice begins with mixing $\sizezi$ number of keys. If Calvin corrupts a link, Alice detects the corrupted link through passive feedback and stops using that link from the next round. After Calvin corrupts $\sizezoo$ links, for any more link that he corrupts, Alice reduces the number of random keys by one. In essence, Alice does not send any data on (up to) $\sizezo$ links that Calvin corrupts, and uses $\sizezii$ number of random keys. Roughly speaking, the scheme is secure because Calvin does not observe anything on $\sizezio$ links, and $\sizezii$ number of random keys protect the data on $\sizezii$ links that Calvin can only eavesdrop.

For additive jamming, Alice leverages the fact that Calvin cannot observe data on $\sizezoo$ links. Here, when Calvin corrupts a link, she starts sending random symbols on that link. The idea is that, even after Calvin corrupts the symbols by adding any noise of his choice, $\sizezoo$ of them can be used as keys for the next round. This limits the number of explicit keys to be mixed and enhances the rate.

The following theorem formally states the capacity expressions for the above settings. 
\begin{theorem}[Secrecy capacity with Causal Feedback, Additive Jamming] {\normalfont When causal passive feedback available to the encoder, the capacity for simultaneous reliability and secrecy is given by 
\begin{align*}
\Rjfadd &=
\left\{\begin{array}{ll} (C - z_{r})^{+} & \mbox{if} \: z_r \geq z_w\\
(C - z_{w})^{+} &\mbox{if} \: z_w > z_r \end{array}\right.\\
\end{align*}}
\end{theorem}

\begin{theorem}[Secrecy capacity with Causal Feedback, Overwrite Jamming] {\normalfont When causal passive feedback available to the encoder, the capacity for simultaneous reliability and secrecy is given by
\begin{align*}
\Rjfow = (C - z_{ro} - z_{wo} - z_{rw})^{+}.\\
\end{align*}}
\end{theorem}

As earlier, we give the full proof in Appendix~\ref{sec:appendix_feedback_secrecy}.

\bibliographystyle{IEEEtran}
\bibliography{main_function}

\newpage
\appendices

\section{Auxiliary Lemmas}
\label{sec:lemma}

\begin{definition}[Matrix-hashing]\label{def:hash} 
For any N $\times$ N square matrix $\mbox{A}$ and vector $\vec{\rho}$ of length N, the matrix-hash $h(A, \vec{\rho})$ is defined as $\vec{\theta} = A\vec{\rho}$, where $\vec{\theta}$ is also of length N. 
\end{definition}

\begin{lemma}\label{lem:1} 
Let $q$ be a prime power. Suppose the $\N \times \N$ matrix $A$ is uniformly distributed over $\Fq^{\N \times \N}$ and the length-$\N$ non-zero vector $\vec{\rho}$ is also uniformly distributed over $F_q^N\setminus\{0\}$. Let $\hat{A}$ be a random matrix independent from $A$ and $\vec{\rho}$ be distributed over $\Fq^{\N \times \N}$ with some distribution $P_{\hat{A}}(\cdot)$. Then for any $\vec{\rho}$ not equal to zero,
$$\underset{\hat{A}, A, \rho}{\Pr}[\hat{A}\vec{\rho} = A\vec{\rho}] \le \frac{1}{q}.$$
\end{lemma}

\emph{Proof:} The matrices $A$ and $\hat{A}$ are uniformly distributed over finite field $\Fq$, and they are also independent from each other. Therefore the difference of the two matrices $\hat{A} - A$ should also be a matrix which is uniformly distributed over $\Fq^{\N \times \N}$. We use the fact that a random matrix over $\Fq$ is non-invertible with probability at most $\frac{1}{q}$. Since $\vec{\rho}$ is non-zero, the equation $(\hat{A} - A)\vec{\rho} = 0$ only if the matrix $\hat{A} - A$ is non-invertible. Therefore we conclude that $$\underset{\hat{A}, A, \rho}{\Pr}[\hat{A}\vec{\rho} = A\vec{\rho}] = \underset{\hat{A}, A, \rho}{\Pr}[(\hat{A} - A)\vec{\rho} = 0] \le \frac{1}{q}.$$  

\begin{lemma}\label{lem:2} 
Let $q$ be a prime power and suppose the $\N \times \N$ matrix $A$ is uniformly distributed over $\Fq^{\N \times \N}$. Let $\hat{\theta}$ be a random length-$N$ vector distributed over $\Fq^{\N}$ independently from $A$ and according to an arbitrary distribution $P^{\vec{\rho}}_{\hat{\theta}}(\cdot)$. Then for any non-zero length-$N$ vector $\vec{\rho}$,
$$\underset{A, \vec{\hat{\theta}}}{\Pr}[A\vec{\rho} = \vec{\hat{\theta}}] \le \frac{N}{q}.$$
\end{lemma}

\emph{Proof:} Each row $\vec{A}_i$ of matrix $A$ is uniformly distributed over the finite field $\Fq^N$ and independent from vector $\vec{\rho}$ by the definition of matrix $A$. For every $i$ \emph{s.t.} $1 \le i \le N$, $\vec{\theta}_i = \vec{A}_i \vec{\rho}$ is also uniformly distributed over $\Fq$. Therefore, the vector $\vec{\theta}$ is uniformly distributed over $\Fq^N$ and for each $i$, $$\underset{A, \vec{\hat{\theta}}}{\Pr}[\vec{\theta}_i = \vec{\hat{\theta}}_i] \le \frac{1}{q}$$ by Schwartz-Zipple Lemma.
By applying the Union bound, we conclude that $$\underset{A, \vec{\hat{\theta}}}{\Pr}[A\vec{\rho} = \vec{\hat{\theta}}] = \underset{A, \vec{\hat{\theta}}}{\Pr}[\vec{\theta} = \vec{\hat{\theta}}] \le \frac{N}{q}.$$

\section{Proofs for reliablity without feedback}
\label{sec:proof_causal}
In the presence of a causal adversary, a two-part rate-region is presented in Section IV. In this Appendix we provide supporting proofs for Theorems 1 and 2, which provide the claimed characterizations of the two-part rate-regions for additive jamming and overwrite jamming.

\subsection{Proof of Theorem~\ref{thm:additiveequal}}
\subsubsection{Achievability for weak adversary regime}

Since the adversary can only jam in a causal manner, a \emph{pairwise-hashing scheme} is helpful to achieve the best rate for the weak adversary regime.

{\em \underline{Encoder}:} The encoder operates over a finite field ${\Fq}$, where $q = 2^{\bb}$  and we set ${\bb} = \log({\n}{\C})$ for equal link capacities while $\bb = \log({\n}{\chat})$ for unequal link capacities. Choose $\N$ such that $\N^2+\N(\C+1)=\frac{n}{\bb}$. The codeword consists of $\C$ vectors $\vecX_1,\vecX_2,\ldots,\vecX_\C$, each of length $\N^2+\N(\C+1)$ over the finite field $\Fq$. Each $\OneX$ is of the form $$\OneX=\primex.$$

Here, $\OneU$ is such that $\vecU_1,\vecU_2,\ldots,\vecU_\C$ together form the codeword corresponding to the length-n${\Rjaddc}$ bit message $\m$ using a Reed-Solomon code of length $\C$ and rate $\C-\sizezio - \sizezoo$ over the finite field $\Fq^{\N^2}$. $\OneK$ is the hash key generated for each row. The value of hash function $h({\SecondU}, {\OneK})$ is defined as $\vechij$. For each link $\Li$, the encoder also appends $\C$ pairwise hashes $\vec{h_{i1}}, \vec{h_{i2}}, ... , \vec{h_{iC}}$ corresponding to the $\C$ links.

{\em \underline{Hash Function}:} The idea of {\it Matrix-hashing} (see Lemma~\ref{lem:1} in Appendix~\ref{sec:lemma}) is used for the hash function. Notice that the randomly generated hash keys are of size $N$ and each length-${\N}^2$ payload ${\OneU}$ can be rearranged as a $\N \times \N$ matrix $D_i$. We pad auxiliary bits before the packets if a square matrix cannot be formed.
The hashes $\vechij$ is obtained from the hash function $h({\SecondU}, \OneK) = D_j \OneK$.

{\em \underline{Decoder}:} After transmission, the received packet of the $i$-th link $\Li$ is of the form 
$$\OneY = \primey.$$ For each $i, j$, link $\Li$ and $L_j$ are consistent if and only if $\vec{h}_{ij}^\prime = h(\SecondU^\prime, \OneK^\prime)$ and $\vec{h}_{ji}^\prime = h(\OneU^\prime, \SecondK^\prime)$. In particular, link $\Li$ is called self-consistent if and only if $\vec{h}_{ii}^\prime = h(\OneU^\prime, \OneK^\prime)$. 

The decoder Bob first removes the links that belong to $\zoo$ by checking self-consistency. Then Bob constructs an undirected graph with $\C$ vertices and for $\forall i, j$, he connects the two vertices $v_i$ and $v_j$ if $L_i$ and $L_j$ are consistent. To detect the uncorrupted links, Bob adopts a \emph{``finding largest clique''\footnote{The method, comes up in\cite{JagLHE:05}, is not an NP complete problem.}} strategy, i.e., a link is assumed to be uncorrupted if its corresponding vertex belongs to the largest clique. Finally, Bob decodes the message from the $\C - \sizezo$ uncorrupted links by Reed-Solomon code.

{\em \underline{Analysis}:} Before transmission, any two links are consistent and the clique formed by Alice is of size $\C$. After adversarial corruption, the size of clique formed by Alice (\emph{correct clique}) is of size at least $\C - \sizezo$. On the other hand, Calvin may mimic the behavior of Alice and and attempt to form a fake clique that is as large as possible. For any two links belonging to $\zio$, Calvin is able to make them consistent since he can modify the payload first, and then compute matching hashes to insert. However, if link $L_i$ belongs to $\zio$ and link $L_j$ belongs to $\zii$, Calvin cannot induce consistency since $\vec{h}_{ji}^\prime \ne h(\OneU^\prime, \SecondK^\prime)$ with high probability. This is because with causality, Calvin doesn't have the ability to observe $\SecondK$ and $\vec{h}_{ji}$ when modifying $\OneU$. In this scenario, the probability that Calvin can induce $\vec{h}_{ji}^\prime = h(\OneU^\prime, \SecondK^\prime)$ is at most $\frac{1}{q}$ over finite field ${\Fq}$ (see Lemma~\ref{lem:1} in Appendix~\ref{sec:lemma}). We would like to make $q$ larger, \emph{i.e.}, enlarge the size of the packet, to reduce the error probability.

In conclusion, Calvin is able to form a fake clique of size at most $\sizezio$. If Bob wishes to figure out the uncorrupted links by finding largest clique, the size of correct clique should be larger than the fake clique, i.e. $\C - \sizezo > \sizezio$. Therefore we derive the condition for our weak adversary regime, which is $\sizezoo + 2\sizezio < \C$, and the rate $R = \C - \sizezio - \sizezoo$ can be achieved.
\footnote{Notice that the links belong to $\zoo$ are also detectable by simply checking the self-consistency. We claim a link belongs to $\zoo$ if it is not self-consistent. However, this process is not necessary since we can detect the uncorrupted links by finding the largest clique.}

\subsubsection{Converse for weak adversary regime}

Irrespective of the encoding/decoding scheme, Calvin can always add uniformly random noise to any $\sizezo$ links. No information can be recovered from the $\sizezoo$ links and thus no rate higher than $\C - \sizezo$ is possible. 

\subsubsection{Converse for strong adversary regime}

In the strong adversary regime ($\sizezoo + 2\sizezio \ge \C$), we prove that no reliable communication is possible no matter which encoding scheme Alice will use. To confuse Bob, the causal adversary Calvin will always adopt the following \emph{``symmetrization''} strategy: (a) corrupt the last $\sizezoo$ (\emph{resp. } $\sizezoo + 1$) links by adding random noise if $\C - \sizezoo$ is even (\emph{resp.} odd), and (b) ``attack" either the top half or the bottom half of the remaining $\cpri = \C - \sizezoo$ (\emph{resp.} $\cpri = \C - \sizezoo - 1$) links, where the ``attack" is defined below. This is a viable jamming strategy for Calvin since $\sizezio \ge (\C - \sizezoo) / 2$ in the strong adversary regime. The specific attack Calvin chooses is to pick a message $\mpri$ first and substitute the original codewords belong to $\zio$ by the codewords corresponding to $\mpri$. In this way, Bob has no idea whether $\m$ or $\mpri$ was transmitted.

We assume the message {\M} is uniformly distributed from the message set $\mathcal{M}$, denoted by $U_{\M}$. Let $\randomX$ be the random variable of the codeword. Moreover, $\randomMX$ is used to represent the random variable of the codeword conditioned on message $\m$. The event $\Gamma(\m, \key, \mpri, \prikey)$ stands for the scenario when Alice chooses a message $\m$ and a random key $\key$ while Calvin chooses a message $\mpri$ and a random key $\prikey$. We can show that the probability of the event $\Gamma(\m, \key, \mpri, \prikey)$ and the event $\Gamma(\mpri, \prikey, \m, \key)$ are exactly the same.
\begin{align*}
&Pr[\Gamma(m, k, m^\prime, k^\prime)] \\
=&U_{M}(m)P_{X_R(m)}(X)U_{M}(m^\prime)P_{X_R(m^\prime)}(X^\prime) \\
=&U_{M}(m^\prime)P_{X_R(m^\prime)}(X^\prime)U_{M}(m)P_{X_R(m)}(X)  \\
=& Pr[\Gamma(m^\prime, k^\prime, m, k)]
\end{align*}

Let $P_y({\m}, {\key}, {\mpri}, {\prikey})$ and $P_y({\mpri}, {\prikey}, {\m}, {\key})$ denote the distributions of the received codeword conditioned on the events $\Gamma(\m, \key, \mpri, \prikey)$ and $\Gamma(\mpri, \prikey, \m, \key)$ respectively. Given the event $\Gamma(\m, \key, \mpri, \prikey)$, the received codeword would be either 
$$[\overrightarrow{X}_1,..., \overrightarrow{X}_{\frac{C^\prime}{2}}, \overrightarrow{X}^\prime_{\frac{C^\prime}{2} + 1},..., \overrightarrow{X}^\prime_{C^\prime}, \overrightarrow{N}_{C^\prime + 1},..., \overrightarrow{N}_C]$$
or
$$[\overrightarrow{X}^\prime_1,..., \overrightarrow{X}^\prime_{\frac{C^\prime}{2}}, \overrightarrow{X}_{\frac{C^\prime}{2} + 1},..., \overrightarrow{X}_{C^\prime}, \overrightarrow{N}_{C^\prime + 1},..., \overrightarrow{N}_C]$$
with equal probability. The same distribution of the codeword will be received when the event $\Gamma(\mpri, \prikey, \m, \key)$ happens. We conclude that the distributions $P_y({\m}, {\key}, {\mpri}, {\prikey})$ and $P_y({\mpri}, {\prikey}, {\m}, {\key})$ are exactly the same. Therefore Bob cannot distinguish the events $\Gamma(\m, \key, \mpri, \prikey)$ and $\Gamma(\mpri, \prikey, \m, \key)$ when decoding, and thus has no idea whether ${\m}$ or ${\mpri}$ are transmitted. The error probability is 
$$\Pr(error) = \frac{1}{2} (1 - \frac{1}{2^{n{\Rjaddc}}})$$
if Bob uses an optimal maximum-likelihood decoder (the term $1 - \frac{1}{2^{n{\Rjaddc}}}$ is due to the ``small'' probability that the message $m'$ Calvin chooses to use to confuse Bob happens to actually match Alice's message $m$).

\subsection{Proof of Theorem~\ref{thm:overwriteequal}}

\subsubsection{Achievability for the weak adversary regime}

The pairwise-hashing scheme also works for the overwrite jammer. The encoding scheme here is the same as that in the additive case. We briefly describe it below for completeness. As earlier, we first generate a payload $\OneU$ for each link $L_i$ using a Reed-Solomon code. Next, we append the hash key and pairwise hashes to the the payload to obtain the codeword ${\OneX} = {\primex}$. After transmission, the received packet is denoted by $\OneY = {\primey}$. As in the additive case, the decoder forms an undirected decoding graph with $\C$ nodes and connects two vertices $v_i$ and $v_j$ if $L_i$ and $L_j$ are consistent. Finally, the decoder finds the largest clique in the decoding graph to determine the set of uncorrupted links. Although the same encoding strategy is applied, a different rate-regime is obtained since the overwrite jammer is slightly stronger than the additive one.

{\em \underline{Analysis}:} After transmission, the size of the \emph{correct clique} is at least $\C - \sizezo$ since Calvin doesn't have privilege to jam on these links. At the same time, Calvin is able to induce a \emph{fake clique} of size no larger than $\sizezio + \sizezoo$. This is because on the links that belong to $\zio$ and $\zoo$, Calvin can overwrite the payloads first and then overwrite the hash vectors with appropriately computed replacements.. Therefore the corresponding vertices will form a clique of size $\sizezio + \sizezoo$. Notice that from the receiver's perspective, the subset $\zio$ is equivalent to $\zoo$ with overwrite jamming. With additive jamming, we have proved the fake clique that Calvin may induce is of size $\sizezio$. Therefore as long as $\sizezio + \sizezoo < \C - \sizezo$, the rate $R = \C - \sizezo$ is achievable.

\subsubsection{Converse for weak adversary regime}

Irrespective of the encoding/decoding scheme, Calvin can always arbitrarily choose a subset $\zo$ and overwrite these links by adding noises. As a result, at most $\C - \sizezo$ links can carry useful information and thus the maximum rate is at most  $\C - \sizezo$.

\subsubsection{Converse for strong adversary regime}

The condition for strong adversary regime is $\sizezio + \sizezoo \ge \C / 2$ with overwrite jamming. In this regime, the adversary is able to jam at least half of the $\C$ links and may perform a similar \emph{``symmetrization''} strategy. Irrespective of the coding scheme, Calvin will (a) first pick a message $\mpri$ and a key $\prikey$ randomly to obtain the corresponding codeword $\Xpri$, (b) then attack either the top half or the bottom half (If $\C$ is odd, Calvin will first ``erase'' one link by overwrite it with a zero packet). In this case, the messages $\m$ and $\mpri$ are perfectly symmetric so that Bob is unable to distinguish them.

If the event $\Gamma(\m, \key, \mpri, \prikey)$ happens, the received codeword would be either 
$$[\overrightarrow{X}_1,..., \overrightarrow{X}_{\frac{C}{2}}, \overrightarrow{X}^\prime_{\frac{C}{2} + 1},..., \overrightarrow{X}^\prime_{C}]$$
or 
$$[\overrightarrow{X}^\prime_1,..., \overrightarrow{X}^\prime_{\frac{C}{2}}, \overrightarrow{X}_{\frac{C}{2} + 1},..., \overrightarrow{X}_{C}]$$
with equal probability. Meanwhile, the received codeword has the same distribution when the event $\Gamma(\mpri, \prikey, \m, \key)$ happens. Therefore the distributions $P_y({\m}, {\key}, {\mpri}, {\prikey})$ and $P_y({\mpri}, {\prikey}, {\m}, {\key})$ are also the same and Bob is unable to decide which message is transmitted. The error probability is equal to $\Pr(error) = \frac{1}{2}(1 - \frac{1}{2^{n{\Rjowc}}})$ if a random decision is made. 

\section{Proofs for reliablity and secrecy without feedback}
\label{sec:proofs-secrecy-additive}
\subsection{Proof of Theorem~\ref{thm:secret-reliable-rate-additive}}
\subsubsection{Weak adversary Regime}
\label{sec:additive-case-1}
${}$\\
\underline{Converse:} Consider the following strategy for Calvin. First, on the links in $\zo$ he adds uniformly random noise that is independent of the codewords on other links. Next, he eavesdrops on all $\zi$ links. We show that, using standard information-theoretic inequalities, that it is not possible for Alice to reliably and secretly transmit at any rate more than $C - \sizezo - \sizezi$. Notice that Calvin can jam any $\sizezo$ links and can eavesdrop any $\sizezi$ links. Consider a code of length $n$ that achieves an error probability $\epsilon_n$ and achieves perfect secrecy. The following set of inequalities follow.
\begin{IEEEeqnarray}{rCl}
\Hp{\M} & = & \Hcond{\M}{\Y} + \I{\M}{\Y} \nonumber\\
& \stackrel{(a)}{\leq} & n\epsn + \I{\M}{\Y} \nonumber\\
& \stackrel{(b)}{=} & n\epsn + \I{\M}{\Y_{1}^{\sizezo}} + \I{\M}{\Y_{\sizezo+1}^C | \Y_{1}^{\sizezo}} \nonumber\\
& \stackrel{(c)}{\leq} & n\epsn +  \I{\M}{\X_{\sizezo+1}^C} \nonumber\\
& \stackrel{(d)}{\leq} & n\epsn + \I{\M}{\X_{\sizezo+1}^{\sizezo+\sizezi}} + \I{\M}{\X_{\sizezo+\sizezi+1}^C | \X_{\sizezo+1}^{\sizezo+\sizezi}} \nonumber\\
& \stackrel{(e)}{\leq} & n\epsn + \Hcond{\X_{\sizezo+\sizezi+1}^C}{\X_{\sizezo+1}^{\sizezo+\sizezi}}\nonumber\\
& \stackrel{(f)}{\leq} & n\epsn + C - \sizezo - \sizezi,
\end{IEEEeqnarray}
where $\epsn \to 0$ as $n\to\infty$. Here, (a) follows from Fano's inequality. Inequalities (b) and (d) follow from the chain rule for mutual information. To obtain (c), we assume without loss of generality that Calvin jams first $\sizezo$ links. Then, we get $\I{\M}{\Y_{1}^{\sizezo}} =0$, as Calvin adds uniform random noise independent of Alice's transmissions. Also, $\I{\M}{\Y_{\sizezo+1}^C | \Y_{1}^{\sizezo}} = \I{\M}{\Y_{\sizezo+1}^C}$ due to independence of added noise. Finally, we use the fact that for the set of uncorrupted links, we have $Y_{C\setminus\zo} = X_{C\setminus\zo}$, which gives $\I{\M}{\Y_{\sizezo+1}^C} = \I{\M}{\X_{\sizezo+1}^C}$. For getting (e), we use the fact that for any subset $\zi$ of links of size $\sizezi$, the secrecy requirement imposes that $\I{\M}{\X_{\zi}} = 0$. Thus, $\I{\M}{\X_{\sizezo+1}^{\sizezo+\sizezi}} = 0$. In addition, we have $\I{\M}{\X_{\sizezo+\sizezi+1}^C | \X_{\sizezo+1}^{\sizezo+\sizezi}} \leq \Hcond{\X_{\sizezo+\sizezi+1}^C}{\X_{\sizezo+1}^{\sizezo+\sizezi}}$. Lastly, (f) follows from the fact $\Hcond{\X_{\sizezo+\sizezi+1}^C}{\X_{\sizezo+1}^{\sizezo+\sizezi}} \leq \Hp{\X_{\sizezo+\sizezi+1}^C} \leq C-\sizezo-\sizezi$, where the second inequality is due to unit link capacities.  

\underline{Achievability:}
Alice first appends $n(C-\sizezo-\sizezi)$ message symbols with $n\sizezi$ uniform random key symbols to form $n(C-\sizezo)$ {\it super-message} symbols. Then, she uses the achievable scheme mentioned in the proof of Theorem 1 (case 1) composed of a $(\C, \C - \sizezo)$ Reed-Solomon code together with the pairwise hashing scheme. We require that the generator matrix of the Reed-Solomon code consists of a Cauchy matrix. 

Now, Bob can locate the set $\zo$ of corrupted links using pairwise hashing and uses the Reed-Solomon code to decode the super-message symbols from the remaining links. Then, Bob separates the random keys and the message symbols from the super-message symbols, since the first $n(C-\sizezo-\sizezi)$ symbols of the super-message are the message symbols. 

For secrecy, we show that Calvin cannot infer any information from the links he eavesdrops. We denote the set of random keys by $\key$ and the corresponding random variable by $\K$. Further, let $\X_{\zi}$ denote the links being eavesdropped. Consider the following set of inequalities.
\begin{IEEEeqnarray}{rCl}
\I{\M}{\X_{\zi}} & = & \Hp{\X_{\zi}} - \Hcond{\X_{\zi}}{\M} \nonumber\\
& \stackrel{(g)}{\leq} & n\sizezi - \Hcond{\X_{\zi}}{\M} \nonumber\\
& \stackrel{(h)}{\leq} & n\sizezi - \Hcond{\X_{\zi}}{\M} + \Hcond{\X_{\zi}}{\M,\K} \nonumber\\
& \stackrel{(i)}{=} & n\sizezi -  \I{\X_{\zi}}{\K|\M} \nonumber\\
& \stackrel{(j)}{=} & n\sizezi - \Hcond{\K}{\M} + \Hcond{\K}{\M,\X_{\zi}} \nonumber\\
& \stackrel{(k)}{=} & n\sizezi - \Hp{\K} + \Hcond{\K}{\M,\X_{\zi}} \nonumber\\
& \stackrel{(l)}{\leq} & \Hcond{\K}{\M,\X_{\zi}},
\end{IEEEeqnarray}
where (g) follows from from the fact that each link has unit capacity and Calvin can eavesdrop at most $\sizezi$ links, (h) follows from the non-negativity of entropy, (i) and (j) follow from the definition of mutual information, (k) follows because keys are independent of the message, and (l) follows from the fact that the keys are uniform random, giving $\Hp{\K} = n\sizezi$. Now, in order to prove secrecy, we need to show that $\Hcond{\K}{\M,\X_{\zi}} = 0$. In other words, one can decode the keys from $\X_{\zi}$ and $\M$. Let $G_{(\X_{\zi})}$ denote the rows of the Cauchy generator matrix corresponding to the symbols $\X_{\zi}$. Therefore, we have 
$$\X_{\zi} = G_{(\X_{\zi})} \twomatrix{\M}{\K}.$$

To prove that $\Hcond{\K}{\M,\X_{\zi}} = 0$, one needs to show that the following system of linear equations can be solved.
\begin{equation}
\label{eq:secrecy-scheme}
\X_{\zi} = \twomatrix{G_{(\X_{\zi})}}{I \:\: | \:\: O} \twomatrix{\M}{\K},
\end{equation}
where $I$ denotes identity matrix of size $n(C-\sizezo-\sizezi) \times n(C-\sizezo-\sizezi)$, and $O$ denotes zero matrix of size $n(C-\sizezo-\sizezi) \times n\sizezi$. First notice that $G_{(\X_{\zi})}$ is a Cauchy matrix since it is a sub-matrix of a Cauchy matrix. Then, using the property that any square sub-matrix of a Cauchy matrix is non-singular, it is straightforward to show that the matrix $\twomatrix{G_{(\X_{\zi})}}{I \:\: | \:\: O}$ is invertible. Therefore, the linear system~\eqref{eq:secrecy-scheme} can be inverted, and we have $\Hcond{\K}{\M,\X_{\zi}} = 0$.

\subsubsection{Strong adversary Regime}
\label{sec:additive-case-2}

Notice that even in the absence of secrecy, no positive rate is achievable in this regime. Adding the extra requirement of secrecy can only decrease the communication rate. Thus no communication at positive rate is possible in this regime.

\subsection{Proof of Theorem~\ref{thm:secret-reliable-rate-overwrite}}
\label{sec:proofs-secrecy-overwrite}
For the weak adversary case, the converse and achievability proofs follow from the same arguments as in the proof of Theorem 5 and is omitted for brevity.
 
For the strong adversary regime, the rate without the secrecy requirement is zero. This implies that no positive rate is possible when secrecy condition is added on top of reliability.

\section{Proof for reliability with passive feedback}
\label{sec:proof_of_feedback}
\subsection{Proof of Theorem~\ref{thm:ajcf}}
Noting that the rate cannot exceed $\C-\sizezo$ even with feedback (since Calvin can always inject random noise on $\sizezo$ links). Therefore, to prove the claim of Theorem~\ref{thm:ajcf}, it suffices to show the following:
\begin{itemize}
\item (a). $\C-\sizezo$ is achievable whenever $\sizezi<\C$ or $\C>2\sizezo$,
\item (b). No positive rate is possible when $\sizezi=\C$ and $\C \leq 2\sizezo$. 
\end{itemize}

\subsubsection{Proof of (a)} We prove the achievability of the rate $\C-\sizezo$ by partitioning the parameter set $\zaddfb=\{\sizezi<\C\}\cup \{\C>2\sizezo\}$ into disjoint sets ${\cal Z}_1=\{\sizezi<\C\}$ and ${\cal Z}_2=\{\sizezi=\C\}\cap\{\C>2\sizezo\}$, and using different coding schemes in the two sets.

 {\em Achievability for ${\cal Z}_1$:} The code operates over two rounds. In the first round, Alice uses an erasure code of length $\C$ capable of correcting upto $\sizezo$ erasures. Upon observing the codewords received by Bob (through passive feedback), Alice computes random hashes of each of the $\C$ received codewords and sends these hashes on the links which are not corrupted in the first round. 
 
 Formally, we show that the rate $\C-\sizezo$ is achievable (and hence, any smaller rate is also achievable). Alice first chooses a blocklength $n>\lceil\log_2{C}\rceil$ and encodes an $nR$-bit message over $n+C\sqrt{n}$ time slots using a two-round scheme as follows. 
{\noindent\em\underline{Round 1}:} In the first round, Alice uses $n$ time slots to transmit using the following scheme. Alice treats $\m$ as $R$ consecutive symbols $m_1,m_2,\ldots,m_R$ from a finite field $\mathbb{F}_{2^n}$ of size $2^n$. Next, Alice encodes $(m_1,m_2,\ldots,m_R)$ to $\Xone=(\xone_1,\xone_2,\ldots,\xone_\C)$ using a Reed-Solomon code capable of correcting upto $\sizezo$ erasures and sends $\xone_i$ on the link $L_i$ for each $i=1,2,\ldots,\C$. These codewords are corrupted by Calvin and Bob receives $\Yone=(\yone_1,\yone_2,\ldots,\yone_\C)$. Alice also observes $\Yone$ causally using the passive feedback available to her. Based on her observation, Alice partitions $L_1,L_2,\ldots,L_\C$ into two sets -- $\Ztrue$ consisting of all links $L_i$ where $\xone_i=\yone_i$ and $\Zfalse$ consisting of all links $L_i$ where $x_i\neq y_i$. Note that $\Zfalse\subseteq \zo\subsetneq\{\alllink\}$.

{\noindent\em\underline{Round 2}:} In the second round, Alice uses the feedback seen from first round and transmits random hashes over $2C\sqrt{n}$ time slots using the following scheme. Alice picks $\C$ independent random keys $\rho_1,\rho_2,\ldots,\rho_\C$ and computes hashes $h(\yone_1,\rho_1),h(\yone_2,\rho_2)$, $\ldots$, $h(\yone_\C,\rho_\C)$, each of length $\lceil\sqrt{n}\rceil$ using the matrix hash scheme (See Appendix~\ref{sec:lemma}). Next, Alice transmits the codeword $\xtwo_i=[h(\yone_1,\rho_1)\ h(\yone_2,\rho_2)\ \ldots h(\yone_\C,\rho_\C)\ \rho_1\ \rho_2\ \ldots\ \rho_\C]$ on every link $L_i\in \Ztrue$ and a random length-$2C\sqrt{n}$ vector $\xtwo_i$ on every link  $L_i\in \Zfalse$.  

{\noindent\em\underline{Decoding}:} Bob first partitions the set of links into sets $\Zhattrue$ and $\Zhatfalse$ using the following classification. 
\begin{itemize}

\item If Bob determines that the hash values and keys specified by $\ytwo_i$ are consistent with all the received codewords in the first round, {\em i.e.}, $\Yone$, he assigns $L_i$ to $\Zhattrue$. \item Else, Bob assigns $L_i$ to $\Zhatfalse$.
\end{itemize}
Finally, if the size of $\Zhattrue$ is at least as large as $\C-\sizezo$, Bob uses the codewords $(\xone_i:L_i\in\Zhattrue)$ to decode the message. Else, he declares an error.   

{\noindent\em\underline{Analysis}:} Note that $\Zhattrue$ includes every link that is not corrupted by Calvin (and possibly some other links as well). Thus, $|\Zhattrue|\geq\C-\sizezo$. Since the Reed-Solomon erasure correcting code can recover $\Xone$ from any $\C-\sizezo$ correct symbols out of $\xone_1,\xone_2,\ldots,\xone_\C$, in order to prove that Bob can successfully decode $\m$ with a high probability, it is sufficient to show that $\Zfalse\subseteq\Zhatfalse$ with a high probability. Without loss of generality, we show the above when $\sizezo<\C$ (since zero rate is trivially achieved when $\sizezo=\C$).

Let $L_s\in\Zfalse$. Since $\sizezi<\C$, there is at least one link $L_t$ such that $\yone_t$ is observed by Bob (and hence by Alice), but not by Calvin. This implies that in the second round, even if Calvin knows $\rho_t$, he can only randomly guess a consistent replacement for $h(\yone_t,\rho_t)$. Thus,

\begin{eqnarray*}
\lefteqn{\Pr\left(\Zfalse\nsubseteq\Zhatfalse\right)}\\
&\leq&\Pr\left(s\in\Zhattrue\middle\vert s\in\Zfalse \right)\\
&=&\Pr\left(\ytwo_s=[h(\yone_1,\hat{\rho}_1)\ldots h(\yone_\C,\hat{\rho}_\C)\ \hat{\rho}_1\ldots\hat{\rho}_\C]\right.\\
&&\qquad\qquad \mbox{ for some }\hat{\rho}_1,\ldots,\hat{\rho}_\C\Big)\\
&\leq&\Pr_{\xtwo_t,\ytwo_s}\left(\left(\ytwo_s\right)_{\lceil\sqrt{n}\rceil(s-1)+1}^{\lceil\sqrt{n}\rceil s}=h(\yone_t,\hat{\rho}_t)\middle\vert\hat{\rho}_t\right)\\
&\leq&2^{-\sqrt{n}}.
\end{eqnarray*}
In the above, the upper bound on the guessing probability follows from Lemma~\ref{lem:2}.

{\em Achievability for ${\cal Z}_2$:} In this parameter setting, we note that since $\sizezi=\C$, $\zo=\zio$ and $\zoo=\emptyset$. Using the fact that $\sizezo<\C/2$, Theorem~\ref{thm:additiveequal}, we can achieve a rate $\C-\sizezio-\sizezoo$ without using feedback.

\subsubsection{Proof of (b)} Next, we show that $\sizezi=\C$ and Calvin controls more than half the links, Bob cannot distinguish between Alice and Calvin. Let $Z_1=\{L_1,\ldots,L_{\lfloor\C/2\rfloor}\}$ and $Z_2=\{L_{\lceil\C/2\rceil+1},\ldots,L_\C\}$. Let $Z_E=\{L_{\lceil\C/2\rceil}\}$ if $\C$ is odd, and $\emptyset$ otherwise. Note that $|Z_1|=|Z_2|<\sizezo$. 

For any message $\m$ and any code chose by Alice, Calvin's attack strategy is the following. First, Calvin adds a uniformly random noise sequence on any link in $Z_E$. Next, he selects a random message $\m'$ and chooses a set $Z'$ uniformly at random from $\{Z_1,Z_2\}$. Finally, he uses Alice's coding strategy to encode the message $m'$ over the links in $Z'$. Calvin can do this as $\sizezi=\C$ and therefore, anything that is seen by Alice and Bob is also seen by Calvin.

Now, Bob is unable to reliably distinguish between following two cases:\begin{itemize}
\item  Alice's message is $\m$ and Calvin's message is $\m'$.
\item  Alice's message is $\m'$ and Calvin's message is $\m$.
\end{itemize}
Thus, the probability of error for Bob is at least $1/4$ if Alice's rate is non-zero.\qed

\subsection{Proof of Theorem~\ref{thm:ojcf}} From the point of view of the proof of Theorem~\ref{thm:ajcf}, the overwrite jamming case differs from the additive one only in the respect that even on the links in $\zoo$, Calvin knows the output (since he can replace it with anything of his choice). Thus the scheme from the proof of Theorem~\ref{thm:ajcf} works only if there is at least one link which is neither seen nor corrupted by Calvin. 

The converse also follows similar arguments. If Calvin can see the output on all the links, he can follow a symmetrization strategy to confuse Bob if the rate is larger than $\C-2(\sizezio+\sizezoo)$. The idea is that if Calvin sees everything also seen by Bob and Alice, he can follow's Alice's coding scheme exactly but with a different message.

The proof is essentially the same as the proof of Theorem 7, and so we omit it here.

\section{Causal Jamming with Passive Feedback and Secrecy}
\label{sec:appendix_feedback_secrecy}
\subsection{Proof of Theorem 9}
Note that the converse arguments in this case are straightforward. When $\sizezo \geq \sizezi$, the rate cannot exceed $\C - \sizezo$ since Calvin can always inject random noise on $\sizezo$ links. When $\sizezi > \sizezo$, any set of $\sizezi$ links cannot carry meaningful information due to secrecy requirement, which results in the rate $\C-\sizezi$.    

We present the proof of achievability for the regime $\sizezi \geq \sizezo$. The scheme for $\sizezo > \sizezi$ is analogous.
 
{\noindent\em\underline{Encoding}:} The protocol operates over multiple rounds in four stages. Alice divides $nR$-bit message into $NR$ number of symbols, where $N < n$ is such that $N | n$. Observe that each symbol is over the finite field $\GF{q}$ of size $q = 2^{n/N}$. Let $\sizezotilde$, $\sizezotilde \leq \sizezo$, be the number of links that Calvin chooses to corrupt in first $N$ rounds. Let $i_j$ be the round at which Calvin chooses to corrupt $j$-th of the $\sizezo$ links that he can corrupt. Notice that $1\leq\ i_1 \leq i_2 \leq \cdots \leq i_{\sizezotilde} \leq N$. In each round, Alice transmits encoded symbols over $\C$ links as described below.

{\textbf{Stage 1:}} The first stage is from rounds $1 \leq i \leq i_1$. In the first stage, in each round, Alice encodes $R = \C - \sizezi$ message symbols (each over $\GF{2^{n/N}}$) together with $\sizezi$ random keys (chosen uniformly and independently over $\GF{2^{n/N}}$). Let us denote the $R$ message symbols transmitted during $i$-th round as $m^{(i)}_1, m^{(i)}_2, \cdots, m^{(i)}_R$, and the $\sizezi$ random keys as $k^{(i)}_1, k^{(i)}_2, \cdots, k^{(i)}_{\sizezi}$. Then, alice generates $\C$ encoded symbols for the $i$-th round as 
\begin{equation}
\label{eq:stage-1-cw}
\begin{bmatrix} x^{(i)}_1 \\ x^{(i)}_2 \\ \vdots \\ x^{(i)}_{\C}\end{bmatrix} = G \begin{bmatrix} m^{(i)}_1 \\ \vdots \\ m^{(i)}_R \\ k^{(i)}_1 \\ \vdots \\ k^{(i)}_{\sizezi}\end{bmatrix},
\end{equation}
where $G$ is a $\C \times \C$ Cauchy matrix with each entry from $\GF{2^{n/N}}$. 
 
{\textbf{Stage 2:}} The second stage is from rounds $i_1 + 1 \leq i \leq N$. Consider the case when Calvin has corrupted the $j$-th link $(1\leq j \leq \sizezotilde)$. Since Alice can overhear the transmissions till round $i_j$, she knows the links that Calvin has corrupted so far. We denote the set of corrupted links as ${Z_w^{i_j}} = \{{l_1}, \cdots, {l_j}\}$, and its complement as $\bar{Z}_w^{(i_j)}$. Let $y^{(i-1)}_{Z_w^{i_j}} = \{y^{(i-1)}_{l_1}, y^{(i-1)}_{l_2}, \cdots, y^{(i-1)}_{l_j}\}$ be the set of codewords received by Bob in round $i-1$ on the corrupted links. Then, the codewords transmitted in round $i$ $(i_1 + 1 \leq i \leq N)$, in stage 2, are given as
\begin{equation}
\label{eq:stage-2-cw-1}
x^{(i)}_{\goodlinks{j}} = G_{\goodlinks{j}} \begin{bmatrix} m^{(i)}_1 \\ \vdots \\ m^{(i)}_R \\ k^{(i)}_1 \\ \vdots \\ k^{(i)}_{\sizezi - j} \\ y^{(i-1)}_{\badlinks{j}}\end{bmatrix}, 
\end{equation}    
\begin{equation}
\label{eq:stage-2-cw-2}
x^{(i)}_{\badlinks{j}} = \begin{bmatrix} k^{(i)}_{\sizezi - j +1} \\ \vdots \\ k^{(i)}_{\sizezi}\end{bmatrix},\end{equation}    
where $G_{\goodlinks{j}}$ is the sub-matrix of $G$ formed by taking the rows of $G$ corresponding to indices in the set $\goodlinks{j}$.
 
{\textbf{Stage 3:}} Since Alice overhears the symbols received by Bob in each round, Alice can easily determine $i_1, i_2, \cdots, i_{\sizezotilde}$. In stage 3, Alice sends pair of corrupted links and corrupted indices, i.e., $x^{(N+1)}_l = \{i_1, l_1; i_2, l_2; \cdots; i_{\sizezotilde}, l_{\sizezotilde}\}$, on all of the links $1\leq l\leq \C$. 

{\textbf{Stage 4:}} In the last stage, Alice computes a randomized hash of $y_l^{(1:N+1)}$ for each $1\leq l\leq \C$, as follows. First, Alice picks $\C$ independent random keys $\rho_1,\rho_2,\ldots,\rho_\C$ and computes hashes $H = \{h(y_1^{(1:N+1)},\rho_1),h(y_2^{(1:N+1)},\rho_2)$, $\ldots$, $h(y_\C^{(1:N+1)},\rho_\C)\}$, each of length $\lceil\sqrt{n}\rceil$ using the matrix hash scheme (see appendix). Next, Alice transmits the hash $H$ on every link $L_l\in L_{\goodlinks{\sizezotilde}}$ and a random $n$-length vector $k^{(N+2)}_l$ on every link  $L_l\in L_{\badlinks{\sizezotilde}}$.  

{\noindent\em\underline{Decoding}:} The first phase of decoding works in the same way as in the scheme without secrecy. Bob first partitions the set of links into sets $\Zhattrue$ and $\Zhatfalse$ using the following classification.
 
\begin{itemize}
\item For each llink $L_l$, if Bob determines that the hash values and keys specified by $H_l$ are consistent with all the received codewords on that link in the first three stages, {\em i.e.}, $y^{(1:N+1)}$, he assigns $L_l$ to $\Zhattrue$. 
\item Else, Bob assigns $L_l$ to $\Zhatfalse$.
\end{itemize}
From the links in $\Zhattrue$, Bob determines the corrupted links and the rounds from which Bob started corrupting the links, i.e., $\{i_1, l_1; i_2, l_2; \cdots; i_{\sizezotilde}, l_{\sizezotilde}\}$. Bob then decodes the codewords for each round starting from round 1 using the appropriate links for each round. In particular, he uses all the links for rounds $1\leq i\leq i_1 -1$, all the links except $l_1$ for rounds $i_1 + 1\leq i\leq i_2-1$, and so on. If the set $\Zhattrue$ is empty, he declares an error.

{\noindent\em\underline{Analysis for decoding}:} Note that $\Zhattrue$ includes every link that is not corrupted by Calvin. In order to prove that Bob can successfully decode $\m$ with a high probability, first, we need to show that $\badlinks{\sizezo}  = \Zhatfalse$ with a high probability. This proof is analogous to the case without secrecy. 

Next, we need to show that Bob can decode for each round. In stage 1, till rounds $i_1 - 1$, no link is corrupted by Calvin. Thus, Bob can use~\eqref{eq:stage-1-cw} and invert $G$ to decode the messages (and keys as well). In stage 2, let us consider the rounds from $i_j+1 \leq i\leq i_{j+1} - 1$. In each of these rounds, notice that on uncorrupted links $\goodlinks{j}$, we have $y^{(i)}_{\goodlinks{j}} =  x^{(i)}_{\goodlinks{j}}$. Thus, Bob can use~\eqref{eq:stage-2-cw-1} along with the received codewords in previous round on corrupted links $y^{(i-1)}_{\badlinks{j}}$ to get the following system of equations 
\begin{equation}
\label{eq:stage-2-cw-deocde}
\begin{bmatrix} x^{(i)}_{\goodlinks{j}} \\ y^{(i-1)}_{\badlinks{j}} \end{bmatrix} = \begin{bmatrix}G_{\goodlinks{j}} \\ \mathbf{0} \:\: | \:\: I_j \end{bmatrix} \begin{bmatrix} m^{(i)}_1 \\ \vdots \\ m^{(i)}_R \\ k^{(i)}_1 \\ \vdots \\ k^{(i)}_{\sizezi - j} \\ y^{(i-1)}_{\badlinks{j}}\end{bmatrix}, 
\end{equation}    
where $\mathbf{0}$ is a $j\times \C-j$ zero matrix and $I_j$ is a $j\times j$ identity matrix. Using the property of Cauchy matrix that any of its square sub-matrices is non-singular, it is easy to show that one can solve~\eqref{eq:stage-2-cw-deocde}. 

{\textit{Remark:}} Note that in each of the rounds $i_1, i_2, \cdots, i_{\sizezotilde}$, Bob cannot decode the message symbols since Calvin corrupts the new link. Therefore, the number of symbols (messages plus keys) that can be correctly conveyed to Bob is $N - \sizezotilde \geq N - \sizezo \rightarrow N$ for large $N$.

{\noindent\em\underline{Analysis for secrecy}:}  We show that in any round, Calvin does not get any information about the message symbols. In the first stage, in round $i$, $1\leq i\leq i_1$, Calvin gets the following system of equations on the links $Z_r$ that he can observe:
\begin{equation}
\label{eq:stage-1-cw-Calvin}
x^{(i)}_{Z_r} = G_{Z_r} \begin{bmatrix} m^{(i)}_1 \\ \vdots \\ m^{(i)}_R \\ k^{(i)}_1 \\ \vdots \\ k^{(i)}_{\sizezi}\end{bmatrix},
\end{equation}
where $G_{Z_r}$ is the sub-matrix of $G$ rows corresponding to links in $Z_w$. Using the properties of Cauchy matrix, we can prove that $\I{M}{x^{(i)}_{Z_r}} = 0$ using the same steps as in the case without passive feedback. 

Next, we prove secrecy for every round $i$, $i_1 + 1\leq i \leq N$, in stage 2. Consider the case that Calvin has corrupted $j$ links, $1\leq j\leq \sizezotilde$. Let $\sizeziotildej$ be the number of corrupted links that Calvin can also eavesdrop, and $\sizezootildej$ be the remaining corrupted links ($\sizeziotildej+\sizezootildej = j$). Note that Calvin observes (left hand side of) the following system of equations:
\begin{equation}
\label{eq:stage-2-cw-Calvin}
\begin{bmatrix} x^{(i)}_{\sizezi} \\ y^{(i-1)}_{\sizeziotildej} \end{bmatrix} = \begin{bmatrix}G_{\sizezi} \\ \mathbf{0} \:\: | \:\: I_{\sizeziotildej} \end{bmatrix} \begin{bmatrix} m^{(i)}_1 \\ \vdots \\ m^{(i)}_R \\ k^{(i)}_1 \\ \vdots \\ k^{(i)}_{\sizezi - j} \\ y^{(i-1)}_{\sizezootildej} \\ y^{(i-1)}_{\sizeziotildej}\end{bmatrix}, 
\end{equation}    
where $\mathbf{0}$ is a $j\times\sizezootildej$ zero matrix, and $I_{\sizeziotildej}$ is $\sizeziotildej\times\sizeziotildej$ identity matrix. Out of the eavesdropped codewords $x^{(i)}_{\sizezi}$, the ones on the read-write links are random keys~\eqref{eq:stage-2-cw-2}. Thus, in round $i$, the number of codewords containing messages that are observed by Calvin is $\sizezii + \sizezio - \sizezootildej = \sizezi - \sizezootildej$.

Now, observe that Alice mixes $\sizezii - j$ explicit random keys. Out of the randomness generated over $j$ corrupted links by Calvin $\badlinks{j}$ using previous round codewords, Calvin knows $y^{(i-1)}_{\sizeziotildej}$ due to his eavesdropping power on those links. Since he cannot eavesdrop on remaining of the $j$ links, $y^{(i-1)}_{\sizezootildej}$ act as random keys. Thus, the total number of random keys used by Alice are $(\sizezii - j) + (j - \sizeziotildej) = \sizezii - \sizeziotildej$. Hence, the number of keys equals the number of observed codewords. We can prove that no information is leaked using the properties of Cauchy matrix, using the same technique as in the case without passive feedback.

\subsection{Proof of Theorem 10}
The flavor of the scheme and the proof remains the same as above. The key difference is the following. In stage 2, when Calvin corrupts a link, Alice stops using that link and, on the contrary to the additive noise case, she does not initially reduce the number of explicit keys. After Calvin corrupts $\sizezoo$ links, for each additional  corrupted link starting with $(\sizezoo + 1)$-th link, Alice reduces the number of explicit keys by one. Besides, she also stops using that link. Therefore, essentially, Alice does not send any data on (up to) $\sizezo$ links, and on remaining links she mixes message symbols with (at least) $\sizezii$ random keys for secrecy.  

We first present achievability scheme.
 
{\noindent\em\underline{Encoding}:} The protocol operates over multiple rounds in four stages. Alice divides $nR$-bit message into $NR$ number of symbols, where $N < n$ is such that $N | n$. Observe that each symbol is over the finite field $\GF{q}$ of size $q = 2^{n/N}$. Let $\sizezotilde$, $\sizezotilde \leq \sizezo$, be the number of links that Calvin chooses to corrupt in first $N$ rounds. Let $i_j$ be the round at which Calvin chooses to corrupt $j$-th of the $\sizezo$ links that he can corrupt. Notice that $1\leq\ i_1 \leq i_2 \leq \cdots \leq i_{\sizezotilde} \leq N$. In each round, Alice transmits encoded symbols over $\C$ links as described below.

{\textbf{Stage 1:}} The first stage is from rounds $1 \leq i \leq i_1$. In the first stage, in each round, Alice enodes $R = \C - \sizezii - \sizezoo - \sizezio$ message symbols (each over $\GF{2^{n/N}}$) together with $\sizezi$ random keys (chosen uniformly and independently over $\GF{2^{n/N}}$). Let us denote the $R$ message symbols transmitted during $i$-th round as $m^{(i)}_1, m^{(i)}_2, \cdots, m^{(i)}_R$, and the $\sizezi$ random keys as $k^{(i)}_1, k^{(i)}_2, \cdots, k^{(i)}_{\sizezi}$. Then, Alice generates $\C$ encoded symbols for the $i$-th round as follows.
\begin{equation}
\label{eq:ow-stage-1-cw}
\begin{bmatrix} x^{(i)}_1 \\ x^{(i)}_2 \\ \vdots \\ x^{(i)}_{\C}\end{bmatrix} = G \begin{bmatrix} m^{(i)}_1 \\ \vdots \\ m^{(i)}_R \\ k^{(i)}_1 \\ \vdots \\ k^{(i)}_{\sizezi}\end{bmatrix},
\end{equation}
where $G$ is a $\C \times \C - \sizezoo$ Cauchy matrix with each entry from $\GF{2^{n/N}}$. 

{\textbf{Stage 2:}} The second stage is from rounds $i_1 + 1 \leq i \leq N$. Consider the case when Calvin has corrupted the $j$-th link $(1\leq j \leq \sizezotilde)$. Since Alice can overhear the transmissions till round $i_j$, she knows the links that Calvin has corrupted so far. Once Calvin corrupts a link, Alice starts sending random symbols on that link from the next round. Essentially, Alice stops using the corrupted links for any meaningful transmission. If $j\leq\sizezoo$, Alice keeps mixing $\sizezi$ keys, else she reduces the number of keys by one for each corrupted link.

To describe this formally, we denote the set of $j$ links that have been corrupted up to round $i_j$ as ${Z_w^{i_j}} = \{{l_1}, \cdots, {l_j}\}$, and its complement as $\bar{Z}_w^{(i_j)}$. The codewords transmitted in round $i$ $(i_1 + 1 \leq i \leq N)$, in stage 2, are given as
\begin{equation}
\label{eq:ow-stage-2-cw-1}
x^{(i)}_{\goodlinks{j}} = G_{\goodlinks{j}} \begin{bmatrix} m^{(i)}_1 \\ \vdots \\ m^{(i)}_R \\ k^{(i)}_1 \\ \vdots \\ k^{(i)}_{\sizezi - (j - \sizezoo)^{+}} \end{bmatrix}, 
\end{equation}
where $G_{\goodlinks{j}}$ is the sub-matrix of $G$ formed by taking the rows of $G$ corresponding to indices in the set $\goodlinks{j}$ and first $R + \sizezi - (j - \sizezoo)^{+}$ columns, and    
\begin{equation}
\label{eq:stage-2-cw-2}
x^{(i)}_{\badlinks{j}} = \begin{bmatrix} k^{(i)}_{\sizezi - (j - \sizezoo)^{+} +1} \\ \vdots \\ k^{(i)}_{\sizezi - (j - \sizezoo)^{+} + j}\end{bmatrix}.
\end{equation}    

{\textbf{Stage 3:}} Since Alice overhears the symbols received by Bob in each round, Alice can easily determine $i_1, i_2, \cdots, i_{\sizezotilde}$. In stage 3, Alice sends pair of corrupted links and corrupted indices, i.e., $x^{(N+1)}_l = \{i_1, l_1; i_2, l_2; \cdots; i_{\sizezotilde}, l_{\sizezotilde}\}$, on all of the links $1\leq l\leq \C$. 

{\textbf{Stage 4:}} The last stage is similar to the second stage in the scheme for only reliability. The purpose of this stage is to allow Bob to determine the corrupted links and the corresponding rounds. In this stage, Alice computes a randomized hash of the symbols received by Bob in the previous $N$ rounds, i.e., $y_l^{(1:N+1)}$, for each link $1\leq l\leq \C$, as follows. First, Alice picks $\C$ independent random keys $\rho_1,\rho_2,\ldots,\rho_\C$ and computes hashes $H_l = h(y_l^{(1:N+1)},\rho_l$, $1\leq l\leq \C$, each of length $\lceil\sqrt{n}\rceil$ using the matrix hash scheme (see appendix). Next, Alice transmits the hash $H$ on every uncorrupted link $L_l\in L_{\goodlinks{\sizezotilde}}$ and a random $\C\lceil\sqrt{n}\rceil$-length vector $k^{(N+2)}_l$ on every corrupted link  $L_l\in L_{\badlinks{\sizezotilde}}$.  

{\noindent\em\underline{Decoding}:} The first phase of decoding works in the same way as in the scheme without secrecy. Bob first partitions the set of links into sets $\Zhattrue$ and $\Zhatfalse$ using the following classification.
 
\begin{itemize}
\item For each llink $L_l$, if Bob determines that the hash values and keys specified by $H_l$ are consistent with all the received codewords on that link in the first three stages, {\em i.e.}, $y^{(1:N+1)}$, he assigns $L_l$ to $\Zhattrue$. 
\item Else, Bob assigns $L_l$ to $\Zhatfalse$.
\end{itemize}
From the links in $\Zhattrue$, Bob determines the corrupted links and the corresponding rounds from which Bob started corrupting the links, i.e., $\{i_1, l_1; i_2, l_2; \cdots; i_{\sizezotilde}, l_{\sizezotilde}\}$. Bob then decodes the codewords for each round starting from round 1 using the appropriate links for that round. In particular, he uses all the links for rounds $1\leq i\leq i_1 -1$, all the links except $l_1$ for rounds $i_1 + 1\leq i\leq i_2-1$, and so on. If the set $\Zhattrue$ is empty, he declares an error.

{\noindent\em\underline{Analysis for decoding}:} Note that $\Zhattrue$ includes every link that is not corrupted by Calvin. In order to prove that Bob can successfully decode $\m$ with a high probability, first, we need to show that $\badlinks{\sizezo}  = \Zhatfalse$ with a high probability. This proof is analogous to the case without secrecy. 

Next, we need to show that Bob can decode for each round. In stage 1, till rounds $i_1 - 1$, no link is corrupted by Calvin. Thus, Bob can use~\eqref{eq:ow-stage-1-cw} and use any $\C-\sizezoo$ rows of $G$ to decode the messages (and keys as well).\footnote{Here, we use the property of a Cauchy matrix that any of its square sub-matrices is non-singular.} 

In stage 2, let us consider the rounds from $i_j+1 \leq i\leq i_{j+1} - 1$. In each of these rounds, notice that on uncorrupted links $\goodlinks{j}$, we have $y^{(i)}_{\goodlinks{j}} =  x^{(i)}_{\goodlinks{j}}$. Thus, Bob can use~\eqref{eq:ow-stage-2-cw-1} to get the following system of equations 
\begin{equation}
\label{eq:ow-stage-2-cw-1}
x^{(i)}_{\goodlinks{j}} = G_{\goodlinks{j}} \begin{bmatrix} m^{(i)}_1 \\ \vdots \\ m^{(i)}_R \\ k^{(i)}_1 \\ \vdots \\ k^{(i)}_{\sizezi - (j - \sizezoo)^{+}} \end{bmatrix}. 
\end{equation}    
Observe that the number of rows of $G_{\goodlinks{j}}$ is $\C - j$, while the number of columns is $R + \sizezi - (j - \sizezoo)^{+} = \C - \sizezoo - (j - \sizezoo)^{+}$. If $j\leq\sizezoo$, Bob can consider any $\C-\sizezoo$ rows of the Cauchy matrix $G_{\goodlinks{j}}$ to decode for the message and key symbols. If $j>\sizezoo$, $G_{\goodlinks{j}}$ is square and it is non-singular being a Cauchy matrix.

{\textit{Remark:}} Note that in each of the rounds $i_1, i_2, \cdots, i_{\sizezotilde}$, Bob cannot decode the message symbols since Calvin corrupts the new link. Therefore, the number of symbols (messages plus keys) that can be correctly conveyed to Bob is $N - \sizezotilde \geq N - \sizezo \rightarrow N$ for large $N$.

{\noindent\em\underline{Analysis for secrecy}:}  We show that in any round, Calvin does not get any information about the message symbols. In the first stage, in round $i$, $1\leq i\leq i_1$, Calvin gets the following system of equations on the links $Z_r$ that he can observe:
\begin{equation}
\label{eq:stage-1-cw-Calvin}
x^{(i)}_{Z_r} = G_{Z_r} \begin{bmatrix} m^{(i)}_1 \\ \vdots \\ m^{(i)}_R \\ k^{(i)}_1 \\ \vdots \\ k^{(i)}_{\sizezi}\end{bmatrix},
\end{equation}
where $G_{Z_r}$ is the sub-matrix of $G$ rows corresponding to links in $Z_w$. Using the properties of Cauchy matrix, we can prove that $\I{M}{x^{(i)}_{Z_r}} = 0$ using the same steps as in the case for secrecy without passive feedback. 

Next, we prove secrecy for every round $i$, $i_1 + 1\leq i \leq N$, in stage 2. Consider the case that Calvin has corrupted $j$ links, $1\leq j\leq \sizezotilde$. Let $\sizeziotildej$ be the number of corrupted links that Calvin can also eavesdrop, and $\sizezootildej$ be the remaining corrupted links ($\sizeziotildej+\sizezootildej = j$). Note that Calvin observes (left hand side of) the following system of equations:
\begin{equation}
\label{eq:ow-stage-2-cw-1}
\begin{bmatrix} x^{(i)}_{\sizezio-\sizeziotildej} \\ x^{(i)}_{\sizezii} \end{bmatrix} = G_{\sizezio-\sizeziotildej+\sizezii} \begin{bmatrix} m^{(i)}_1 \\ \vdots \\ m^{(i)}_R \\ k^{(i)}_1 \\ \vdots \\ k^{(i)}_{\sizezi - (j - \sizezoo)^{+}} \end{bmatrix}, 
\end{equation}
\begin{equation}
\label{eq:ow-stage-2-cw-Calvin}
x^{(i)}_{\sizeziotildej}  =  \begin{bmatrix}  k^{(i)}_{\sizezi - (j - \sizezoo)^{+} +1} \\ \vdots \\ k^{(i)}_{\sizezi - (j - \sizezoo)^{+} + \sizeziotildej} \end{bmatrix}, 
\end{equation}    
On $\sizeziotildej$ links, Calvin observes random keys. Thus, we need to worry about the remaining of the $\sizezi-\sizeziotildej$ links that he eavesdrops. The number of random key symbols mixed with message symbols is $\sizezi - (j - \sizezoo)^{+}$. If $j\leq\sizezoo$, the number of mixed keys is $\sizezi$, while if $j>\sizezoo$, the number of mixed keys is $\sizezi - (j - \sizezoo) = \sizezi - \sizeziotildej - \sizezootildej + \sizezoo$. Since $\sizezootildej \leq \sizezoo$, it is easy to see that the number of mixed keys is greater than equal to the number of eavesdropped symbols involving messages. Using the properties of Cauchy matrix, one can easily prove that the secrecy requirement is satisfied.

{\noindent\em\underline{Converse}:} Consider a strategy for Calvin wherein he overwrites zero symbols on the $\sizezoo$ links that he can only corrupt. He eavesdrops all the $\sizezi$ links of his choice. We consider that the communication is over multiple rounds, say $t$ number of rounds.
\begin{IEEEeqnarray}{rCl}
\Hp{\M} & = & \Hcond{\M}{\Y^{(1:n)}} + \I{\M}{\Y^{(1:n)}} \nonumber\\
& \stackrel{(a)}{\leq} & n\epsn + \I{\M}{\Y^{(1:n)}} \nonumber\\
& \stackrel{(b)}{=} & n\epsn + \I{\M}{\Y_{\sizezii+\sizezio}^{(1:n)}} + \I{\M}{\Y_{\sizezoo}^{(1:n)} | \Y_{\sizezii+\sizezio}^{(1:n)}} \nonumber\\ 
& \ \ \ +& \I{\M}{\Y_{\C-\sizezoo-\sizezii-\sizezio}^{(1:n)} | \Y_{\sizezoo+\sizezii+\sizezio}^{(1:n)}} \nonumber\\
& \stackrel{(c)}{=} & n\epsn +  \I{\M}{\Y_{\C-\sizezoo-\sizezii-\sizezio}^{(1:n)} | \Y_{\sizezii+\sizezoo+\sizezio}^{(1:n)}} \nonumber\\
& {\leq} & n\epsn +  \Hp{\Y_{\C-\sizezoo-\sizezii-\sizezio}^{(1:n)}} \nonumber\\
& \stackrel{(d)}{\leq} & n\epsn + n(C - \sizezoo - \sizezii - \sizezio), \nonumber
\end{IEEEeqnarray}
where $\epsn \to 0$ as $n\to\infty$. Here, (a) follows from Fano's inequality, (b) follows from the chain rule of mutual information. To obtain (c), first note that $\I{\M}{\Y_{\sizezii+\sizezio}^{(1:n)}} \leq \I{\M}{\X_{\sizezii+\sizezio}^{(1:n)}}$ by data processing inequality. But, for secrecy, we need $\I{\M}{\X_{\sizezii+\sizezio}^{(1:n)}} = 0$. Also, $\I{\M}{\Y_{\sizezoo}^{(1:n)} | \Y_{\sizezii+\sizezio}^{(1:n)}} = 0$ since $\Y_{\sizezoo}^{(1:n)} = 0$ due to Calvin's attack strategy. Finally, (d) follows from unit link capacities.

\end{document}